\documentclass[conference]{IEEEtran}

\usepackage{comment}
\usepackage{amsmath,amssymb,amsfonts}
\usepackage{algorithm}
\usepackage{pgfplots}
\pgfplotsset{compat=1.11,
    /pgfplots/ybar legend/.style={
    /pgfplots/legend image code/.code={%
       \draw[##1,/tikz/.cd,yshift=-0.25em]
        (0cm,0cm) rectangle (3pt,0.8em);},
   },
}
\usepackage[noend]{algpseudocode}
\usepackage{booktabs}
\usepackage{csquotes}
\usepackage{comment}
\usepackage{graphicx}
\usepackage{textcomp}
\usepackage{multirow}
\usepackage{url}
\usepackage{hyperref}
\usepackage{graphicx}
\usepackage{enumitem}

\definecolor{custmaroon}{HTML}{800000}
\definecolor{custbrown}{HTML}{9a6324}
\definecolor{custteal}{HTML}{469990}
\definecolor{custnavy}{HTML}{000075}
\definecolor{custred}{HTML}{e6194b}
\definecolor{custorange}{HTML}{f58231}
\definecolor{custyellow}{HTML}{ffe119}
\definecolor{custgreen}{HTML}{3cb44b}
\definecolor{custcyan}{HTML}{42d4f4}
\definecolor{custblue}{HTML}{4363d8}
\definecolor{custmagenta}{HTML}{f032e6}
\definecolor{custgrey}{HTML}{a9a9a9}
\definecolor{custpink}{HTML}{fabed4}
\definecolor{custbeige}{HTML}{fffac8}
\definecolor{custmint}{HTML}{aaffc3}
\definecolor{custlavender}{HTML}{dcbeff}

\newcommand{\para}{\noindent\textbf}
\newcommand{\X}{\mathbb{X}}
\newcommand{\R}{\mathbb{R}}

\begin{document}

\title{Fast Topology-Aware Lossy Data Compression with Full Preservation of Critical Points and Local Order}

\author{\IEEEauthorblockN{Alex Fallin\IEEEauthorrefmark{1},
Nathaniel Gorski\IEEEauthorrefmark{2},
Tripti Agarwal\IEEEauthorrefmark{2},
Bei Wang\IEEEauthorrefmark{2},
Ganesh Gopalakrishnan\IEEEauthorrefmark{2},
Martin Burtscher\IEEEauthorrefmark{1}}

\IEEEauthorblockA{\IEEEauthorrefmark{1}Texas State University, USA}
\IEEEauthorblockA{\IEEEauthorrefmark{2}University of Utah, USA}

Emails: \{waf13, burtscher\}@txstate.edu, 
\{gorski, beiwang\}@sci.utah.edu, 
\{tripti.agarwal@, ganesh@cs.\}utah.edu
}

\maketitle % should come after the abstract

\begin{abstract}
Many scientific codes and instruments generate large amounts of floating-point data at high rates that must be compressed before they can be stored. Typically, only lossy compression algorithms deliver high-enough compression ratios. However, many of them provide only point-wise error bounds and do not preserve topological aspects of the data such as the relative magnitude of neighboring points. Even topology-preserving compressors tend to merely preserve some critical points and are generally slow. Our Local-Order-Preserving Compressor is the first to preserve the full local order (and thus all critical points), runs orders of magnitude faster than prior topology-preserving compressors, yields higher compression ratios than lossless compressors, and produces bit-for-bit the same output on CPUs and GPUs.
\end{abstract}


\section{Introduction}
\label{sec:intro}
Many scientific instruments and simulations generate data volumes that far exceed what can be practically managed, both in terms of throughput and storage capacity~\cite{scientific_size}. To address this challenge, two primary compression strategies are employed: lossless and lossy. Lossless compression reproduces the original data exactly but often fails to achieve the high compression ratios required for large-scale scientific workflows. In contrast, lossy compression can provide substantially higher compression ratios depending on the chosen error bound, at the cost of discarding some information, making perfect reconstruction of the original data impossible.

%When processing scientific data, 
Topological data analysis (TDA) employs structures such as critical points, contour trees~\cite{carr2003computing}, and Morse-Smale complexes~\cite{edelsbrunner2001hierarchical} to describe, summarize, and analyze complex data. These topological descriptors are derived from \emph{global} properties of the data, making the \emph{local} guarantees provided by standard scientific data compressors insufficient for their preservation. To overcome this limitation, a variety of compressors and frameworks have been developed to preserve various topological aspects of scalar fields~\cite{gorski2025general, li2024msz, soler2018topologically, yan2023toposz}, vector fields~\cite{liang2022toward, xia2025tspsz}, and tensor fields~\cite{gorski2025tfz}.

Unfortunately, existing topology-preserving compressors suffer from several drawbacks. First, they tend to be relatively slow: the preservation of topological information often requires substantial computation, taking orders of magnitude longer than conventional compression methods~\cite{yan2023toposz}.  Second, most approaches preserve only limited topological information, such as the critical points on the contour tree~\cite{yan2023toposz}. Third, many lossy compressors claim error control but do not strictly guarantee the error bound~\cite{codac}.

Our Local-Order-Preserving Compressor (LOPC) addresses these challenges by providing a strict error-bound guarantee, incorporating a new algorithm that preserves the full local order (and thus all critical points), and leveraging parallelism-friendly CPU and GPU implementations to significantly accelerate computation.

Whereas the topology of the data---defined here as the relationships among critical points---is inherently a global property, our approach focuses on preserving the relative ordering of neighboring points, referred to as the \emph{local order}. Maintaining the local order serves as a foundational step toward preserving global topology: it improves the accuracy of local gradient flows, reduces visual artifacts such as jagged structures, and yields more accurate derived quantities.

Critical points of scalar fields (namely local minima, maxima, and saddles) are fundamental topological features that are determined entirely by the local order of the data. They serve as essential descriptors in visualization and form the basis for more complex topological constructs, such as the contour tree and the Morse-Smale complex. Moreover, critical points often carry direct physical meaning in a wide range of domains, including chemistry~\cite{bhatia2018topoms}, climate science~\cite{li2025tracking}, and physics~\cite{sousbie2011persistent}. 
Because local order is defined solely from local information, it is an ideal property to preserve under lossy compression. Our method perfectly preserves the local order and, by extension, all critical points. To the best of our knowledge, no prior lossy data compressor fully preserves either property.

Since LOPC preserves more information, it compresses less than the other tested topology-preserving compressors. However, it is faster even serially. When run in parallel, it reaches speeds that are orders of magnitude higher than the related work, all while preserving local ordering, critical points, and a guaranteed error bound, a unique set of attributes among compressors in the literature.

Our main contributions include:

\begin{itemize}[leftmargin=*]
\item \textbf{A local-order-preserving compression algorithm}: LOPC is the first algorithm that guarantees full preservation of the local order and works in combination with a guaranteed error-bounded lossy compressor. % and includes parallel implementations for both CPUs and GPUs.
\item \textbf{Theoretical guarantees}: We show that LOPC preserves all critical-point locations and types (despite the lossy nature of the underlying compressor), introduces no spurious critical points, and terminates.
\item \textbf{Implementation and optimization}: We provide a detailed description of the implementation, optimization, and parallelization strategies used in LOPC for CPUs and GPUs.
\item \textbf{Comprehensive evaluation}: We compare LOPC to 8 state-of-the-art compressors, 3 of which are topology-preserving, demonstrating its superior effectiveness in preserving critical points and local order.
\end{itemize}
%\begin{itemize}
%    \item A local-order-preserving compression algorithm, LOPC, that operates in tandem with a guaranteed-error-bounded lossy data compressor. It includes parallel CPU and GPU implementations.
%    \item Proofs that LOPC retains all critical-point locations and types (despite the lossy nature of the main compressor), does not introduce new critical points, and terminates.
%    %\item A detailed description of the implementation, optimization, and parallelization of LOPC for CPUs and GPUs.
%    \item A detailed comparison of LOPC with 6 other compressors from the literature demonstrating the effectiveness of LOPC compared to the state of the art.
%\end{itemize}

Upon conclusion of the anonymization phase, our LOPC C++/OpenMP and CUDA codes will be open-sourced. 
The rest of this paper is organized as follows.
Section~\ref{sec:background} provides background. Section~\ref{sec:related} summarizes related work. Section~\ref{sec:appr} explains the LOPC algorithm. Section~\ref{sec:methodology} describes the evaluation methodology. Section~\ref{sec:res} presents and discusses the results. Section~\ref{sec:conc_and_future} concludes the paper with a summary.

\section{Background}
\label{sec:background}

\para{Piecewise-Linear Scalar Fields}: LOPC targets piecewise-linear (PL) scalar functions defined on triangular (2D) or tetrahedral (3D) meshes. Let $\X$ be a triangular mesh and $f:\X \rightarrow \R$ a PL function. Then, the value of $f$ is explicitly stored for each vertex $v$ of $\X$ and extended to all of $\X$ by linear interpolation. Let $\sigma$ be a triangular cell of $\X$ with vertices $v_1$, $v_2$, and $v_3$. Then, any $x \in \sigma$ can be written uniquely in terms of its barycentric coordinates $x = t_1v_1 + t_2v_2 + t_3v_3$, where $t_i \geq 0$ and $\sum_i t_i = 1$. We set $f(x) = \sum_i t_if(v_i)$, which also holds for the analogous definition in 3D. LOPC works with scalar data defined on regular 2D and 3D grids, which it subdivides into triangular and tetrahedral meshes, respectively, as is done in prior work~\cite{Vidal2021Progressive}.

\para{Critical Points}: The critical points of a PL function $f:\X \rightarrow \R$ are defined as follows. Let $v \in \X$ be a vertex. The \textit{link} of $v$, $Lk(v)$, is the set of vertices of $\X$ adjacent to $v$. The \textit{lower link} of $v$ is the set $Lk^-(v) = \{v' \in Lk(v) : f(v') < f(v)\}$. Similarly, the upper link is the set $Lk^+(v) = \{v' \in Lk(v) : f(v') > f(v)\}$. If $Lk^-(v) = \emptyset$, then $v$ is a local maximum. If $Lk^+(v) = \emptyset$, then $v$ is a local maximum. If $LK^-(v)$ and $Lk^+(v)$ are both nonempty and contain vertices belonging to a single connected component each, then $v$ is an ordinary (non-critical) point. Otherwise, $v$ is a saddle point. This strategy assumes that neighboring points have different function values. We guarantee that they do using Simulation of Simplicity \cite{edelsbrunner1990simulation}.

\section{Related Work}
\label{sec:related}

This section describes
%the 4 non-topology-preserving and 
the %topology-preserving 
state-of-the-art floating-point compressors with which we compare LOPC. Note that all lossless compressors necessarily preserve the full topology.

\subsection{Lossless Compressors}

FPzip~\cite{lindstrom2006fast} is a CPU-based library that supports both lossy and lossless compression of scientific data. It exploits floating-point data coherency to predict values in the input, computes the residuals, stores the data as integers, and uses a fast entropy encoder to achieve not only high compression ratios but also fast compression and decompression. 

Zstandard~\cite{collet2016zstandard} (ZSTD) is a parallel CPU compressor that is based on LZ77~\cite{lz77}, ANS~\cite{duda2009asymmetricnumeralsystems}, and Huffman~\cite{huffman} coding. Unlike many of the other tested compressors, ZSTD is a general-purpose compressors and does not specifically target floating-point data.

FPCompress~\cite{mb-asplos25} is a CPU/GPU parallel lossless floating-point compressor that targets smooth scientific data. It offers two versions for both single- and double-precision floating-point data, one that targets compression speed and one that targets compression ratio. Both speed versions use delta encoding, transformation from two's complement to magnitude sign, and leading bit elimination. The single-precision ratio version uses the same delta encoding and two's complement transformation stage but follows it with a bit shuffling step and repeated zero-byte elimination as the final step. The double-precision ratio version is similar, but a finite context method (FCM) predictor is added before the common compression pipeline. The FCM stage internally increases the data size by a factor of 2, but similar to our splitting of the bins and the subbins (see below), this increase in data size is to enable greater compression down the line. For the purpose of this work, we compare to the GPU versions of both the speed and ratio algorithms from FPCompress.

\subsection{Non-topology Preserving Lossy Compressors}

% We compare to two versions of the lossy compressor SZ. 
% SZ3~\cite{sz3-3,sz3-1,sz3-2} is similar to its predecessor SZ2~\cite{sz2} but is an improvement that generally produces better compression ratios with similar throughput. It uses Lorenzo prediction~\cite{lorenzo} and dynamic spline interpolation. It also adopts entropy coding plus lossless compression after the lossy stage (e.g., Huffman~\cite{huffman} followed by GZIP~\cite{gzip} or ZSTD~\cite{zstd}). SZ3 is a CPU-only compressor. cuSZp~\cite{cuszp} is a GPU-based lossy compressor. It splits the data into blocks and then quantizes and predicts the values in all nonzero blocks, which are ultimately compressed by a fixed-length encoder. The fixed-length encoding is implemented using a bit-shuffle operation. Similar to the SZ compressors, we also use quantization as a lossy step, and our lossless stages also rearrange the data but utilizing different transformations. While both LOPC and the SZ compressors are lossy, the SZ compressors do not maintain the topological information from the original file.

SZ3~\cite{sz3-3,sz3-1,sz3-2} is similar to its predecessor SZ2~\cite{sz2} but generally produces better compression ratios with about the same throughput. It uses Lorenzo prediction~\cite{lorenzo} and dynamic spline interpolation. It also adopts entropy coding plus lossless compression after the lossy stage (e.g., Huffman~\cite{huffman} followed by GZIP~\cite{gzip} or ZSTD~\cite{zstd}). SZ3 is a CPU-only compressor and guarantees that the user-requested error bound will not be violated. We both directly and indirectly compare to SZ3, as it is the base compressor used in TopoA (described below). Similar to the SZ compressors, we also use quantization as a lossy step, and our lossless stages also rearrange the data but utilizing different transformations. While both LOPC and the SZ compressors are lossy, the SZ compressors do not maintain the topological information from the original file. %This is why TopoA is needed to augment SZ3.

% ZFP~\cite{zfpnew,zfp} is a widely used tranform-based compressor. It is specifically designed for in-memory array compression and supports on-the-fly random-access decompression. ZFP splits the input into blocks, converts each value into an integer, performs decorrelation, reorders the data, and converts the values from twos-complement to negabinary representation. Then, it groups the bits from most to least significant. Finally, the shuffled bits are losslessly compressed.
% Our approach only has a few commonalities with ZFP (e.g., converting to negabinary format).

PFPL~\cite{mb-ipdps25a} is a high-throughput CPU and GPU compatible lossy compressor. It first quantizes the input and converts the bin values to magnitude-sign representation, but stores outliers inline rather than separately for performance reasons. It then, like LOPC, splits the input into 16kB chunks to allow for fast parallel compression. Each chunk is compressed using a pipeline of lossless data transformations. First, the data is delta encoded~\cite{delta_modulation} and converted to negabinary. Second, the data is bit shuffled, similar to how it is done in ZFP. 
Last, the data is compressed using repeated zero-byte elimination. % that compresses any repeated zeros that were introduced in the preceding stages. 
Like SZ3 and LOPC, PFPL guarantees that the user-requested error bound will not be violated.

\subsection{Topology Preserving Lossy Compressors}

Several topology preserving lossy compressors for scalar field data exist. The oldest such compressor is TopoQZ~\cite{soler2018topologically}. It takes a single persistence parameter and focuses on preserving critical point pairs with a high persistence~\cite{tierny2012generalized}. The decompressed data contains all pairs of critical points with a persistence above the threshold but no other critical points. Moreover, the locations of the preserved critical points may shift. Unlike TopoQZ, LOPC preserves all critical points as well as their locations. However, TopoQZ preserves persistence relationships, which LOPC does not.

TopoSZ~\cite{yan2023toposz} modifies SZ 1.4~\cite{sz1} to preserve the contour tree~\cite{carr2003computing} of the original data after persistence simplification~\cite{tierny2012generalized}. It losslessly stores each critical point in the contour tree. For each other point, it computes upper and lower bounds based on the contour tree. It then iteratively tightens these upper and lower bounds until the contour tree of the decompressed data matches the ground truth. LOPC focuses on preserving all critical points, not just those stored in the contour tree. However, it does not ensure that the internal connections of the contour tree are preserved.

% mSZ~\cite{li2024msz} is a parallelized topology-preserving compressor. It is focused specifically on Morse-Smale segmentation, which is a topological segmentation that differs from the contour-tree and critical-point preservation discussed above. It augments both SZ3 and ZFP with a similar iterative identify-and-fix operation as LOPC uses.

TopoA~\cite{gorski2025general} is a framework designed to augment existing lossy compressors. It achieves the same goal as TopoSZ and uses a similar approach but employs a progressive strategy to tighten the upper and lower bounds. This bypasses iteratively recomputing the contour tree and boosts performance. It also introduces a more efficient quantization scheme to achieve improved compression ratios. We compare to the TopoA-augmented version of SZ3 because it is the best compressing of the augmented guaranteed-error lossy compressors.

\section{LOPC Algorithm and Implementation}
\label{sec:appr}

LOPC incorporates a modular approach to preserve the critical points and local ordering. % while also maintaining the compressibility of the data. 
% Unlike many of the other topology-aware compressors that augment an existing compressor, LOPC is a full application that applies the error bounding, topology preservation, and compression all at once.

\subsection{Quantization}
\label{sec:lossy_quant}

The first step in LOPC is the lossy quantization. This is where the error bounding of the data occurs. LOPC supports both the point-wise absolute (ABS) error bound and the point-wise normalized absolute error bound (NOA). This means that, for an ABS error bound of $\varepsilon$, each value $x$ must satisfy $|x_{original} - x_{reconstructed}| \leq \varepsilon$. The normalized absolute error bound is the ABS error normalized by the value range $R = x_{max} - x_{min}$, that is, the range between the largest and the smallest value in the input. It is commonly found in other topology preserving codes.

The quantizer uses the supplied error bound $\varepsilon$ to map each floating-point value to a bin. This is accomplished by multiplying the value by $1/\varepsilon$ and rounding the result to the nearest integer, which yields the bin number. Normal ABS quantization uses bins that are twice as large and decodes to the center of the bin, which is never more than $\pm\varepsilon$ from the original value. We must halve the bin size to accommodate the later intra-bin adjustments of the reconstructed values (see below) to maintain the local ordering of the input.

In addition to the ABS quantization bins, LOPC also utilizes a set of ``subbins'' to adjust the values up or down in their bin range without violating the error bound. For example, with an ABS error bound of 0.1, the quantizer maps all values between 0.95 and 1.05 to bin number 10. Without subbins, all of those values would be reconstructed to 1.0. With the subbins, they are strategically reconstructed to some value between 0.95 and 1.05, all of which lie within the prescribed error bound, such that the critical-point locations and types of the input are retained as explained next.

\subsection{Preserving Critical Points and Local Order}
\label{sec:fix-up}

Rather than explicitly preserving critical points, LOPC iteratively corrects all values that violate the less-than relationship (i.e., local order) with their neighbors that have the same bin number. It suffices to fix only values that map to the same bin because the local ordering of values mapped to different bins is automatically preserved, as the quantization function is \emph{monotonic increasing}. Notably, preserving local order is a stronger condition than preserving critical points; it guarantees not only the existence but also the exact types of critical points.
This is accomplished using an iterative approach. Algorithm~\ref{alg:main_loop} shows the overall operation of LOPC, and Algorithm~\ref{alg:subbincomp} outlines the operations of a single iteration of the local-order preservation step.

\begin{algorithm}[!htbp]
\caption{Main LOPC Operation}
\label{alg:main_loop}
\begin{algorithmic}[1]
\For{each point $p$} \Comment{parallel loop}
  %\State idx $\gets$ index of $p$;
  \State $p_{bin} \gets$ ABS or NOA quantized input of $p_{val}$;
  \State $p_{subbin} \gets$ 0;
  \State $p_{flags} \gets$ 0;
\EndFor
\For{each point $p$} \Comment{parallel loop}
%  \State idx $\gets$ index of $p$;
  % \State temp\_flag; \Comment{accumulates neighbor info}
  \For{each neighbor $n$ of $p$}
    % \State temp\_flag $\gets$ n is in same bin as p;
    \State $p_{flags} \gets p_{flags}$ $\cup$ ($n_{bin} = p_{bin}$); %\Comment{same bin num}
    % \State temp\_flag $\gets$ n $<$ p; \Comment{tie breaker on index}
    \State $p_{flags} \gets p_{flags}$ $\cup$ ($n$ $<$ $p$); \Comment{with tie breaker}
  \EndFor
  % \State flags[idx] $\gets$ temp\_flag;
\EndFor
\State worklist1 $\gets$ all input points;
\While {worklist1 is not empty}
  \State worklist2 $\gets$ empty;
  \State Algorithm 2;
  \State swap(worklist1, worklist2);
\EndWhile
\end{algorithmic}
\end{algorithm}

\begin{algorithm}[!htbp]
\caption{Parallel Subbin Computation (one iteration)}
\label{alg:subbincomp}
\begin{algorithmic}[1]
\For{each point $p$ in worklist1} \Comment{parallel loop}
%  \State idx $\gets$ index of $p$;
  \State n\_max $\gets$ 0;
  \For{each same-bin neighbor $n$ of $p$} \Comment{using $p_{flags}$}
    \If {$n$ should be less than $p$} \Comment{using $p_{flags}$}
%      \State n\_idx $\gets$ index of $n$;
      \State tie $\gets$ ($n_{idx} > p_{idx}$);  \Comment{tie breaker: 0 or 1}
      \State val $\gets$ atomicRead($n_{subbin}$);
      \State n\_max $\gets$ max(n\_max, val + tie);
    \EndIf
  \EndFor
  \If {atomicMax($p_{subbin}$, n\_max) $<$ n\_max}
    \State worklist2 $\gets$ worklist2 $\cup$ \{$p$'s greater same-bin neighbors\};
  \EndIf
\EndFor
\end{algorithmic}
\end{algorithm}

At the outset, the subbins are zero and the original data has been binned. Then, LOPC computes a set of flags for each point $p$ in the input. For each neighbor of $p$, the flags record whether the neighbor has the same bin number. If so, the flags further record whether the neighbor's value is less than $p$'s value. A deterministic tiebreaker is used to guarantee that a point is always greater or less than any neighboring point, never equal. %This computation is only performed once. 
The flags represent the ground truth that the final values must match in terms of local order.

Next, the iterative process starts. In each iteration, every point $p$ checks its same-bin neighbors $n$ that should be less than $p$, which are identified quickly by checking the flags. If they all meet the condition, nothing is done. Otherwise, $p$'s subbin value is set to the highest such neighbor's subbin value (or that value plus 1 if the tie breaker is in favor of the neighbor) to establish the correct less-than relationship. Note that subbin values are never decreased. This process repeats until no more changes are made, in which case it terminates.

In the end, the subbin values are as low as possible while guaranteeing the same less-than relationships in the reconstructed data as in the original data. Typically, most subbins end up with small integer values near zero, which is important for compressibility.

\subsection{Compression}
\label{sec:lossless}

The above computation converts each input value into two values, a bin number and a subbin number, effectively doubling the original data size. The result is two arrays holding very different information. First is the quantized bin numbers, which contain the information required to reconstruct the original data values within the requested error bound. Second is the subbin numbers, which contain the information required to recreate the original local ordering.

The information density of these two arrays varies greatly depending on the chosen error bound. For example, if the error bound is small, neighboring values tend to be mapped to distinct bins. Hence, the first array contains most of the information of the original data whereas the second array contains mostly zeros as the subbins are not needed. However, if the error bound is large, neighboring values tend to be mapped to the same bin number. In this case, the first array contains little information whereas the second array contains most of the local ordering of the data.

To boost the compression ratio, we use distinct compression algorithms for the bins and the subbins that are customized to compress the two types of data well. LOPC uses SLEEK's quantizer~\cite{mb-ipdps26-temp} to compute the bin numbers, which allows for guaranteed binning within the error bound without special handling of outliers. To losslessly compress the resulting bin numbers, LOPC employs the lossless portion of PFPL's compression algorithm~\cite{mb-ipdps25a, pfpl-git}. SLEEK and PFPL are effective, fast, and support both CPU and GPU execution.
%DIFFNB\_4 BIT\_4 RZE\_1 (Same as PFPL)
%DIFFNB\_8 BIT\_8 RZE\_1 (double)

\begin{figure*}[!htbp]
\centerline{\includegraphics[width=.95\textwidth]{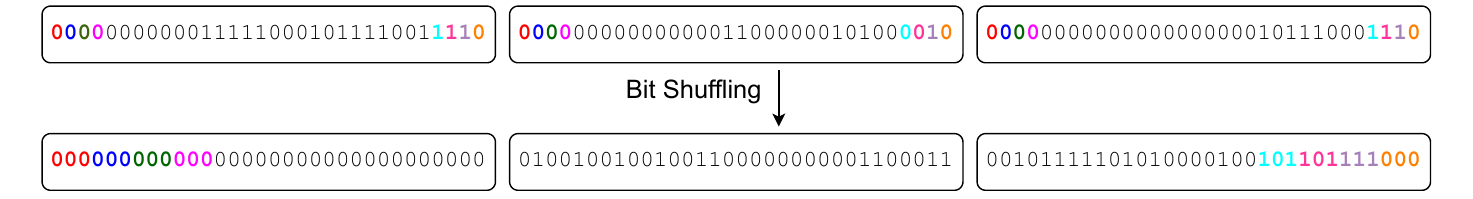}}
\vspace{-1.2mm}
\caption{Example of bit-shuffling lossless stage; for larger inputs, the sequences of bits with the same color are longer}
\label{fig:bitshuffle}
\end{figure*}

\begin{figure*}[!htbp]
\centerline{\includegraphics[width=\textwidth]{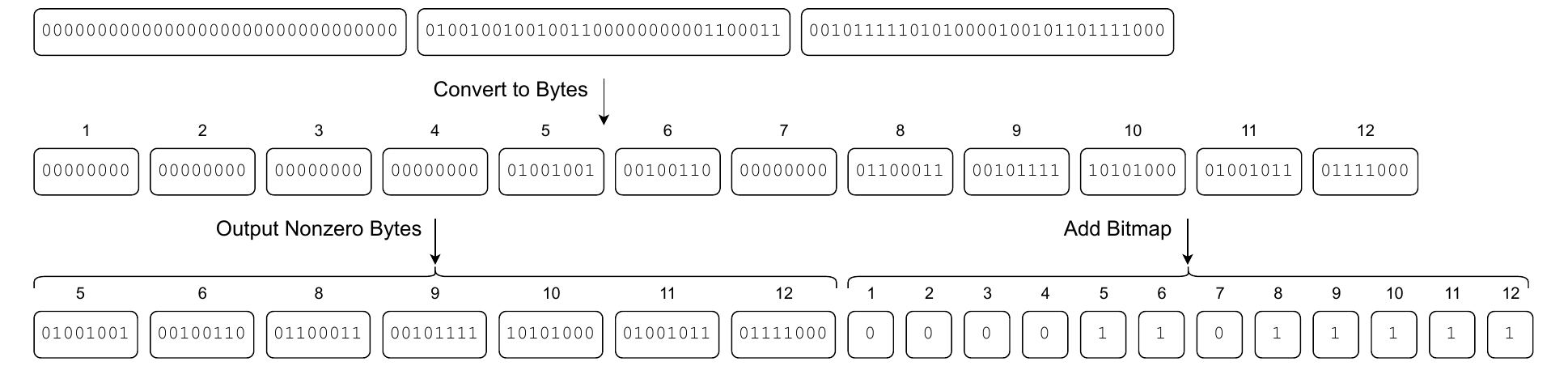}}
\vspace{-1mm}
\caption{Example of zero-byte elimination lossless stage; further compression of the bitmap is not shown}
\label{fig:rze}
\vspace{-1mm}
\end{figure*}

To create good CPU- and GPU-parallel lossless compressors and decompressors for the subbin data, we used the LC framework~\cite{lc-git}. For 32-bit subbins (single-precision data), it generated the 3-stage algorithm BIT\_4 RZE\_4 RZE\_1.
For 64-bit subbins (double-precision data), it generated the 3-stage algorithm BIT\_8 RZE\_8 RZE\_1. In both cases, the first two stages match the word size and the final stage operates at byte granularity. The BIT stages, whose operation is outlined in Figure~\ref{fig:bitshuffle}, perform a bit transposition (or bit shuffle). They group the first bit of every value together, then all the second bits, and so on. The RZE (Repeated Zero Elimination) stages, illustrated in Figure~\ref{fig:rze}, generate a bitmap in which each bit corresponds to a word in the input and indicates whether the word is zero. All zero words are then removed. The compressed output consists of the non-zero words from the input and the bitmap, which itself is repeatedly compressed with a similar algorithm that identifies repeating words rather than zero words. More detail on these stages can be found elsewhere~\cite{mb-asplos25}.

\subsection{Parallelization and Optimization}
\label{sec:parallelization}

LOPC uses already parallel CPU and GPU code from PFPL and SLEEK for computing, compressing, and decompressing the bin information. Moreover, it uses LC-generated CPU- and GPU-parallel lossless compressors and decompressors for the subbin data.

The subbin decoder is embarrassingly parallel as every bin/subbin pair can be independently processed to reconstruct the floating-point value. The parallelization of the subbin encoder is more involved. The flag computation (see Section~\ref{sec:fix-up}) and the subbin initialization are embarrassingly parallel. The iterative phase that raises the subbin values is parallelized as follows. Each iteration runs in a separate barrier interval and increases subbins using an atomicMax operation, making the implementation lock-free. On the GPU, we assign each point to a separate thread. On the CPU, we use a blocked assignment of points to threads.

Since most points require little, if any, adjustment, many subbins reach their final value in the first few iterations. For this reason, processing every point in later iterations is often inefficient. As a remedy, LOPC employs a worklist on which it stores only the greater neighbors of a point whose subbin has just been raised so that only those points will be processed in the next iteration. Moreover, we employ two worklists, one that is read and another that is written. At the end of each iteration, we zero out the size of the old worklist and swap the pointers to the two worklists and their sizes. When filling the worklist, we use an atomicAdd to both increment the worklist size and obtain a unique slot for writing the new element. To avoid duplicates on the worklist, for each point, we record and atomically update the most recent iteration in which it was placed on the worklist.

\subsection{Correctness and Termination}

This subsection explains how the LOPC algorithm preserves local ordering, including all critical points, while guaranteeing termination and error boundedness. Since our quantization is an increasing function, the local order is always correct between neighbors that are quantized to distinct bin numbers. Hence, in the rest of the explanation, we only consider same-bin neighbors. %, i.e., the points that are quantized into the same bin during quantization.

Initially, all subbin numbers are zero. In each iteration, the algorithm examines, for every point, all neighbors that belong to the same bin and whose value should be lower. If a point fails the less-than relationship with any of these neighbors, the subbin number of that point is raised according to one of two rules. (1) The subbin number is raised to match that of the highest violating neighbor if that neighbor's index is lower. (2) The subbin number is raised one higher than that of the highest violating neighbor if that neighbor's index is higher. These two rules ensure that the local ordering among neighboring points gradually becomes consistent (even in the presence of ties) with the local order of the original data.

%As we repeat the process, the local inconsistencies are eliminated. 
The algorithm terminates when no violation remains, i.e., all local relationships are satisfied, thereby fully preserving the local order. Since preserving local order implies preserving local maxima, minima, and saddle points, this in turn guarantees that all critical points and their types are preserved as well.

LOPC is guaranteed to terminate because the update process is non-decreasing and bounded. In each iteration, if a violation occurs, at least one subbin number increases. This monotonicity ensures continuous progress and prevents oscillations (i.e., livelock). Moreover, the highest number a subbin can reach is finite. To see why, it helps to consider the connected component (CC) of same-bin values to which a point belongs.
%values are limited within each connected bin component because the algorithm always follows two properties of the update rule. 
(1) Each such CC must contain at least one local minimum (due to the tie breaker), and the subbin number of a local minimum is never raised because it has no lower same-bin neighbors. (2) No point ever raises its subbin number to more than one higher than its highest same-bin neighbor. (3) The targeted less-than relationships are necessarily acyclic since they stem from the original input data.
Together, these properties limit the possible range of subbin numbers in a CC with $n$ points to between $0$ and $n-1$. %and hence the algorithm cannot increase indefinitely. Consequently, in a connected same bin component containing $n$ points, the highest possible subbin number is $n - 1$. 
Hence, after a finite number of updates, all violations disappear and the process terminates. In the worst case, when all $n$ points form an increasing chain, the algorithm converges after $O(n^2)$ iterations since we can create the values $0$ through $n-1$ with $n \times (n - 1) / 2 = O(n^2)$ individual increments.

The only remaining concern is whether the subbin range provides enough resolution to represent all necessary ordering distinctions while remaining within the quantization bin, thereby guaranteeing error boundedness. This is also guaranteed because the maximum number of subbin levels required to preserve local order within any CC is directly tied to the number of distinct floating-point values present in this CC in the original data. Note that decompression maps each value within its quantization bin such that subbin 0 decodes to the lowest representable value within the bin, subbin 1 to the next lowest, and so on. Thus, the algorithm guarantees the local order without running out of subbin range and without violating the user-provided error bound.

\section{Experimental Methodology}
\label{sec:methodology}

%System 1 is Ithaca, System 2 is Bulach
We compare LOPC to the compressors described in Section~\ref{sec:related} on the two systems listed in Table~\ref{tab:systems}. We ran the GPU codes on System 1 and the CPU codes on System 2. %The operating system for System 1 is Fedora 37. The operating system for System 2 is Fedora 36.
%System 1 is based on an AMD Ryzen Threadripper 2950X CPU with 16 cores. Hyperthreading is enabled, that is, the 16 cores can simultaneously run 32 threads. The main memory has a capacity of 64 GB. The operating system is Fedora 37. The GPU in this system is an NVIDIA RTX 4090 (Ada Lovelace architecture) with 16,384 processing elements distributed over 128 multiprocessors. Its global memory has a capacity of 24 GB. The GPU driver version is 525.85.05.
%System 2 is based on two Intel Xeon Gold 6226R CPUs with 16 cores each. Hyperthreading is also enabled, meaning the 32 cores can run 64 simultaneous threads. The main memory has a capacity of 64 GB. The operating system is Fedora 36. The GPU in this system is an NVIDIA A100 (Ampere architecture) with 6,912 processing elements distributed over 108 multiprocessors. It has 40 GB of global memory. The GPU driver version is 535.113.01.
%The CPU codes were compiled using {\em gcc/g++} version 12.2.1 with the build processes supplied by their respective authors. % with respect to optimization flags.

We compiled the CPU codes using the build processes supplied by their respective authors. When not specified, we used the ``-O3 -march=native'' flags. Unless automatically determined, the thread count was set to the number of CPU cores as hyperthreading usually does not help.
%The GPU codes were compiled with {\em nvcc} version 12.0.140 using the ``-O3 -arch=sm\_89'' flags for the RTX 4090 and the ``-O3 -arch=sm\_80'' flags for the A100. When not specified by the build process, we compiled the C++ codes using the ``-O3 -march=native'' flags.
We compiled the GPU codes using ``-O3 -arch=sm\_89'' for the RTX 4090.
%and the ``-O3 -arch=sm\_80'' flags for the A100. 

For all compressors, we measured the execution time of the compression and decompression functions, excluding reading the input file, verifying the results, and transferring data to and from the device.
%For compression quality, we bounded the maximum point-wise absolute error, maximum point-wise relative error, or maximum point-wise normalized absolute error. %, or PSNR.
We ran each experiment 9 times and collected the compression ratio, median compression throughput, and median decompression throughput as well as the critical-point false positives, false negatives, and false types. 
% The plots report the geometric mean of the geometric mean of each suite so as not to overemphasize suites with more files. Additionally, the use of the geometric rather than arithmetic mean helps dampen any inputs that significantly outperform the general case~\cite{how-not-to-lie-with-statistics}.
For all compressors, we used NOA error bounds of 1E-2 and 1E-4. For TopoA augmented SZ3, we ran two experiments, one where the persistence threshold $\epsilon$ is $1.5\times$ and another where it is $0.5\times$ the NOA error bound. These settings reflect a normal level of persistence and an over-preserving level, respectively.
%We report the results in tables. 
If a compressor runs for more than an hour (wall-clock time), we report `TO' (timeout). If a compressor crashes, fails, or runs out of memory, we report `DNF' (did not finish).

\begin{table}[!htbp]
    \centering
    \caption{Systems used for experiments}
    \vspace{-1mm}
    \resizebox{0.9\columnwidth}{!}{
    \begin{tabular}{|l|r|r|}
    \hline
     & System 1 & System 2 \\ \hline
    \hline
    CPU & %Ryzen
    Threadripper 2950X & Threadripper 3970X \\ \hline
    Base Clock & 3.5 GHz & 3.7 GHz \\ \hline
    Sockets & 1 & 1\\ \hline
    Cores Per Socket & 16 & 32 \\ \hline
    Threads Per Core & 2 & 2 \\ \hline
    %NUMA nodes & 1 & 2\\ \hline
    %Cache line size & 64 bytes \\ \hline
    Main memory & 64 GB & 256 GB \\ \hline
    \hline
    GPU & RTX 4090 & N/A\\ \hline
    Compute Capability & 8.9 & N/A \\ \hline
    Base Clock & 2.2 GHz & N/A \\ \hline
    Boost Clock & 2.5 GHz & N/A \\ \hline
    SMs & 128 & N/A\\ \hline
    %Max Threads per SM & 1024 & 1024 \\ \hline
    CUDA Cores per SM & 128 & N/A\\ \hline
    Main memory & 24 GB HBM2e & N/A \\ \hline
    \hline
    Operating System & Fedora 37 & Ubuntu 24.04.1 LTS\\ \hline
    \texttt{g++} Version & 12.2.1 & 13.3.0 \\ \hline
    \texttt{nvcc} Version & 12.0 & N/A \\ \hline
    GPU Driver & 525.85 & N/A \\ \hline
    \end{tabular}}
    \label{tab:systems}
\end{table}
%\vspace{-4mm}
\begin{table}[hbtp]
    \begin{center}
        \caption{Information about the used inputs}
        \vspace{-1mm}
        \label{tab:inputs}
        \resizebox{1\columnwidth}{!}{
        \begin{tabular} { |l|l|c|c|c|c| }\hline
            \textbf{Name} &\textbf{Description}  &\textbf{Format} &\textbf{Dimensions} &\textbf{Size (MB)}\\\hline
                Isabel & Weather Sim. & Single & 90 $\times$ 500 $\times$ 500 & 90 \\\hline
                %NYX & Cosmology & Single & 512 $\times$ 512 $\times$ 512 & 537 \\\hline
                Tangaroa & Weather Sim. & Single & 300 $\times$ 180 $\times$ 120 & 26 \\\hline
                Earthquake & TeraShake 2 Sim. & Double & 375 $\times$ 188 $\times$ 50 & 28 \\\hline
                Ionization & Ionization Sim. & Double & 310 $\times$ 128 $\times$ 128 & 41 \\\hline
                Miranda & Hydrodynamics & Double & 384 $\times$ 384 $\times$ 256 & 302 \\\hline
                S3D & Weather Sim. & Double & 500 $\times$ 500 $\times$ 500 & 1000 \\\hline
                SCALE-LETKF & LETKF & Double & 1200 $\times$ 1200 $\times$ 98 & 1129 \\\hline
                QMCPACK & Quantum MC & Double & 69 $\times$ 69 $\times$ 115 & 4 \\\hline
        \end{tabular}}
    \end{center}
\end{table}

We used the 2 single- and 6 double-precision inputs listed in Table~\ref{tab:inputs} as inputs for the compressors. The Earthquake dataset is from the TeraShake 2 earthquake simulation~\cite{earthquake_repo,earthquake}. The Ionization dataset is timestep 125 from cluster 2 of an ionization front simulation~\cite{ionization}. The Tangaroa dataset is the wind velocity field from a simulation of the Tangaroa research vessel~\cite{Tangaroa}. The remaining inputs are sourced from the SDRBench repository~\cite{sdrbench_url,sdrbench}, which hosts real-world scientific datasets from various domains for compression evaluation. The table lists the input name, a short description, the data type, input dimensions, and file size. We chose these inputs because they are commonly used as benchmarks in topology work~\cite{gorski2025general,yan2023toposz} and represent the kind of scientific data that may have important topological information to preserve.

\section{Results}
\label{sec:res}
In this section, we evaluate the performance of the compressors discussed in Section~\ref{sec:related} on the inputs described in Section~\ref{sec:methodology}. We first discuss the preservation of critical points and local ordering for the tested compressors. Next, we compare the compression ratios achieved by each compressor using NOA error bounds. Then, we analyze the compression and decompression speed. Next, we study the effect of error bounding on the compression quality and speed. Finally, we evaluate the reconstruction quality yielded by the tested compressors.

\subsection{Critical-Point and Local-Order Preservation}
\label{subsec:crit_points}
\begin{table*}[hbtp]
    \begin{center}
        \caption{Count of false positives, false negatives, and false types (`M' = million, `K' = thousand).}
        \vspace{-1mm}
        \label{tab:fp-fn-ft}
        \resizebox{1.5\columnwidth}{!}{
        \begin{tabular}{lccccccc}
\multicolumn{1}{l|}{}           & \multicolumn{1}{c|}{}      & \multicolumn{2}{c|}{TopoA SZ3}                                  & \multicolumn{1}{l|}{}             & \multicolumn{1}{l|}{}             & \multicolumn{1}{l|}{}             & \multicolumn{1}{l}{}     \\
\multicolumn{1}{l|}{}           & \multicolumn{1}{c|}{LOPC}  & \multicolumn{1}{c}{$\epsilon$ = 1.5x EB} & \multicolumn{1}{c|}{$\epsilon$ = 0.5x EB}    & \multicolumn{1}{c|}{TopoSZ}       & \multicolumn{1}{c|}{TopoQZ}       & \multicolumn{1}{c|}{SZ3}          & \multicolumn{1}{c}{PFPL} \\ \hline
\multicolumn{8}{|c|}{EB = 1E-2}                                                                                                                                                                                                                                       \\ \hline
\multicolumn{1}{l|}{Isabel}     & \multicolumn{1}{c|}{0/0/0} & 686K/16K/2K                  & \multicolumn{1}{c|}{734K/13K/3K} & \multicolumn{1}{c|}{2M/16K/3K}    & \multicolumn{1}{c|}{461K/20K/639} & \multicolumn{1}{c|}{28K/22K/493}  & 430K/22K/314             \\
\multicolumn{1}{l|}{Tangaroa}   & \multicolumn{1}{c|}{0/0/0} & 380K/2K/565                  & \multicolumn{1}{c|}{514K/2K/500} & \multicolumn{1}{c|}{DNF}          & \multicolumn{1}{c|}{57K/4K/236}   & \multicolumn{1}{c|}{15K/5K/711}   & 35K/7K/421               \\
\multicolumn{1}{l|}{Earthquake} & \multicolumn{1}{c|}{0/0/0} & 218K/95K/8K                  & \multicolumn{1}{c|}{180K/83K/8K} & \multicolumn{1}{c|}{448K/95K/11K} & \multicolumn{1}{c|}{22K/100K/1K}  & \multicolumn{1}{c|}{25K/110K/2K}  & 10K/112K/398             \\
\multicolumn{1}{l|}{Ionization} & \multicolumn{1}{c|}{0/0/0} & 95K/12K/1K                   & \multicolumn{1}{c|}{110K/10K/1K} & \multicolumn{1}{c|}{DNF}          & \multicolumn{1}{c|}{13K/15K/753}  & \multicolumn{1}{c|}{23K/18K/2K}   & 16K/20K/1K               \\
\multicolumn{1}{l|}{Miranda}    & \multicolumn{1}{c|}{0/0/0} & 1M/12M/214K                  & \multicolumn{1}{c|}{TO}          & \multicolumn{1}{c|}{644K/13M/255} & \multicolumn{1}{c|}{566/13M/3}    & \multicolumn{1}{c|}{81K/13M/9K}   & 2K/13M/2                 \\
\multicolumn{1}{l|}{S3D}        & \multicolumn{1}{c|}{0/0/0} & 7M/23K/5K                    & \multicolumn{1}{c|}{TO}          & \multicolumn{1}{c|}{133K/50K/218} & \multicolumn{1}{c|}{2M/26K/2K}    & \multicolumn{1}{c|}{83K/40K/4K}   & 2M/46K/2K                \\
\multicolumn{1}{l|}{SCALE}      & \multicolumn{1}{c|}{0/0/0} & 4M/216K/36K                  & \multicolumn{1}{c|}{TO}          & \multicolumn{1}{c|}{8/315K/0}     & \multicolumn{1}{c|}{456K/258K/5K} & \multicolumn{1}{c|}{268K/303K/6K} & 2M/310K/2K               \\
\multicolumn{1}{l|}{QMCPACK}    & \multicolumn{1}{c|}{0/0/0} & 22K/168/34                   & \multicolumn{1}{c|}{15K/163/33}  & \multicolumn{1}{c|}{85K/313/78}   & \multicolumn{1}{c|}{11K/167/36}   & \multicolumn{1}{c|}{2K/575/16}    & 6K/584/5                 \\ \hline
\multicolumn{8}{|c|}{EB = 1E-4}                                                                                                                                                                                                                                       \\ \hline
\multicolumn{1}{l|}{Isabel}     & \multicolumn{1}{c|}{0/0/0} & 3K/2K/162                    & \multicolumn{1}{c|}{3K/1K/132}   & \multicolumn{1}{c|}{TO}           & \multicolumn{1}{c|}{5K/4K/103}    & \multicolumn{1}{c|}{8K/5K/557}    & 14K/8K/305               \\
\multicolumn{1}{l|}{Tangaroa}   & \multicolumn{1}{c|}{0/0/0} & 12K/135/9                    & \multicolumn{1}{c|}{14K/128/8}   & \multicolumn{1}{c|}{70K/203/22}   & \multicolumn{1}{c|}{8K/292/15}    & \multicolumn{1}{c|}{7K/356/30}    & 17K/540/41               \\
\multicolumn{1}{l|}{Earthquake} & \multicolumn{1}{c|}{0/0/0} & 183K/36K/9K                  & \multicolumn{1}{c|}{TO}          & \multicolumn{1}{c|}{199K/43K/11K} & \multicolumn{1}{c|}{50K/41K/4K}   & \multicolumn{1}{c|}{196K/52K/13K} & 78K/60K/6K               \\
\multicolumn{1}{l|}{Ionization} & \multicolumn{1}{c|}{0/0/0} & 87K/4K/1K                    & \multicolumn{1}{c|}{118K/4K/1K}  & \multicolumn{1}{c|}{458K/4K/1K}   & \multicolumn{1}{c|}{10K/7K/814}   & \multicolumn{1}{c|}{53K/6K/1K}    & 10K/8K/864               \\
\multicolumn{1}{l|}{Miranda}    & \multicolumn{1}{c|}{0/0/0} & 624K/13M/74K                 & \multicolumn{1}{c|}{TO}          & \multicolumn{1}{c|}{1M/12M/779K}  & \multicolumn{1}{c|}{7/13M/0}      & \multicolumn{1}{c|}{215K/13M/22K} & 25/13M/0                 \\
\multicolumn{1}{l|}{S3D}        & \multicolumn{1}{c|}{0/0/0} & 27K/2K/204                   & \multicolumn{1}{c|}{TO}          & \multicolumn{1}{c|}{TO}           & \multicolumn{1}{c|}{53K/5K/295}   & \multicolumn{1}{c|}{7K/5K/409}    & 252K/10K/715             \\
\multicolumn{1}{l|}{SCALE}      & \multicolumn{1}{c|}{0/0/0} & 0/0/0                        & \multicolumn{1}{c|}{TO}          & \multicolumn{1}{c|}{TO}           & \multicolumn{1}{c|}{84K/129K/957} & \multicolumn{1}{c|}{219K/155K/9K} & 210K/187K/3K             \\
\multicolumn{1}{l|}{QMCPACK}    & \multicolumn{1}{c|}{0/0/0} & 122/84/16                    & \multicolumn{1}{c|}{120/84/16}   & \multicolumn{1}{c|}{268/138/25}   & \multicolumn{1}{c|}{316/119/34}   & \multicolumn{1}{c|}{135/159/25}   & 1K/341/35               
\end{tabular}
}
    \end{center}
\end{table*}

Table~\ref{tab:fp-fn-ft} shows the quality at which the critical points are preserved. The key strength of LOPC is immediately obvious: it preserves all critical points and local ordering whereas none of the other compressors do, not even the topology-preserving compressors. While TopoA and TopoSZ preserve the critical points on the contour tree, and TopoQZ preserves the critical point pairs, our results show the generally high number of false positives, false negatives, and false types that the other topology-preserving compressors introduce.

Comparing the other topology-preserving compressors to the non-topology-preserving lossy compressors makes it clear that a large portion of the critical points are missed when only preserving the contour tree or critical point pairs. LOPC avoids introducing any false positives, false negatives, or false types by preserving the full local ordering.

The results in the table further demonstrate that it is not straightforward for other topology-preserving compressors to preserve more critical points by lowering the persistence threshold. In many cases, doing so actually introduces more erroneous critical points in a given category. Additionally, lowering the persistence threshold introduces significant additional runtime, leading to timeouts. LOPC avoids these issues without the need for a tunable parameter.

\subsection{Compression Ratios}

\begin{table}[hbtp]
    \begin{center}
        \caption{Comparison with topology-preserving compressors for the 1E-2 NOA error bound}
        \vspace{-1mm}
        \label{tab:1e-2res}
        \resizebox{1\columnwidth}{!}{
        \begin{tabular}{lrrrrrrr}
\multicolumn{1}{l|}{}           & \multicolumn{3}{c|}{LOPC}                                                                 & \multicolumn{2}{c|}{TopoA SZ3}                                                       & \multicolumn{1}{c|}{}                    & \multicolumn{1}{c}{}       \\
\multicolumn{1}{l|}{}           & \multicolumn{1}{c}{Ser} & \multicolumn{1}{c}{OMP} & \multicolumn{1}{c|}{CUDA}             & \multicolumn{1}{c}{$\epsilon$ = 1.5x EB} & \multicolumn{1}{c|}{$\epsilon$ = 0.5x EB} & \multicolumn{1}{c|}{TopoSZ}              & \multicolumn{1}{c}{TopoQZ} \\ \hline
\multicolumn{8}{|c|}{Compression Ratio}                                                                                                                                                                                                                                                    \\ \hline
\multicolumn{1}{l|}{Isabel}     & 5.47                    & 5.47                    & \multicolumn{1}{r|}{5.47}             & \textbf{64.40}                           & \multicolumn{1}{r|}{28.43}                & \multicolumn{1}{r|}{27.50}               & 4.67                       \\
\multicolumn{1}{l|}{Tangaroa}   & 4.39                    & 4.39                    & \multicolumn{1}{r|}{4.39}             & \textbf{28.36}                           & \multicolumn{1}{r|}{20.91}                & \multicolumn{1}{r|}{DNF}                 & 3.33                       \\
\multicolumn{1}{l|}{Earthquake} & 8.79                    & 8.79                    & \multicolumn{1}{r|}{8.79}             & \textbf{85.44}                           & \multicolumn{1}{r|}{25.09}                & \multicolumn{1}{r|}{65.85}               & 7.08                       \\
\multicolumn{1}{l|}{Ionization} & 10.19                   & 10.19                   & \multicolumn{1}{r|}{10.19}            & \textbf{93.45}                           & \multicolumn{1}{r|}{62.69}                & \multicolumn{1}{r|}{DNF}                 & 5.78                       \\
\multicolumn{1}{l|}{Miranda}    & 10.05                   & 10.05                   & \multicolumn{1}{r|}{10.05}            & \textbf{239.09}                          & \multicolumn{1}{r|}{TO}                   & \multicolumn{1}{r|}{92.01}               & 9.14                       \\
\multicolumn{1}{l|}{S3D}        & 9.48                    & 9.48                    & \multicolumn{1}{r|}{9.48}             & 35.90                                    & \multicolumn{1}{r|}{TO}                   & \multicolumn{1}{r|}{\textbf{11,336.58}}  & 5.82                       \\
\multicolumn{1}{l|}{SCALE}      & 10.74                   & 10.74                   & \multicolumn{1}{r|}{10.74}            & 37.16                                    & \multicolumn{1}{r|}{TO}                   & \multicolumn{1}{r|}{\textbf{401,765.12}} & 5.82                       \\
\multicolumn{1}{l|}{QMCPACK}    & 9.07                    & 9.07                    & \multicolumn{1}{r|}{9.07}             & \textbf{102.04}                          & \multicolumn{1}{r|}{82.48}                & \multicolumn{1}{r|}{12.01}               & 9.08                       \\
\multicolumn{1}{l|}{Geomean}    & 8.18                    & 8.18                    & \multicolumn{1}{r|}{8.18}             & 68.32                                    & \multicolumn{1}{r|}{37.80}                & \multicolumn{1}{r|}{\textbf{457.04}}     & 6.04                       \\ \hline
\multicolumn{8}{|c|}{Compression Throughput   (MB/s)}                                                                                                                                                                                                                                      \\ \hline
\multicolumn{1}{l|}{Isabel}     & 9                       & 17                      & \multicolumn{1}{r|}{\textbf{750}}     & 1                                        & \multicolumn{1}{r|}{1}                    & \multicolumn{1}{r|}{0}                   & 14                         \\
\multicolumn{1}{l|}{Tangaroa}   & 1                       & 1                       & \multicolumn{1}{r|}{\textbf{104}}     & 1                                        & \multicolumn{1}{r|}{1}                    & \multicolumn{1}{r|}{DNF}                 & 14                         \\
\multicolumn{1}{l|}{Earthquake} & 7                       & 13                      & \multicolumn{1}{r|}{\textbf{940}}     & 2                                        & \multicolumn{1}{r|}{1}                    & \multicolumn{1}{r|}{1}                   & 21                         \\
\multicolumn{1}{l|}{Ionization} & 3                       & 4                       & \multicolumn{1}{r|}{\textbf{313}}     & 1                                        & \multicolumn{1}{r|}{1}                    & \multicolumn{1}{r|}{DNF}                 & 11                         \\
\multicolumn{1}{l|}{Miranda}    & 5                       & 8                       & \multicolumn{1}{r|}{\textbf{414}}     & 0                                        & \multicolumn{1}{r|}{TO}                   & \multicolumn{1}{r|}{1}                   & 8                          \\
\multicolumn{1}{l|}{S3D}        & TO                      & 9                       & \multicolumn{1}{r|}{\textbf{408}}     & 2                                        & \multicolumn{1}{r|}{TO}                   & \multicolumn{1}{r|}{2}                   & 24                         \\
\multicolumn{1}{l|}{SCALE}      & TO                      & TO                      & \multicolumn{1}{r|}{34}               & 1                                        & \multicolumn{1}{r|}{TO}                   & \multicolumn{1}{r|}{2}                   & \textbf{34}                \\
\multicolumn{1}{l|}{QMCPACK}    & 12                      & 16                      & \multicolumn{1}{r|}{\textbf{730}}     & 1                                        & \multicolumn{1}{r|}{0}                    & \multicolumn{1}{r|}{0}                   & 13                         \\
\multicolumn{1}{l|}{Geomean}    & 5                       & 8                       & \multicolumn{1}{r|}{\textbf{314}}     & 1                                        & \multicolumn{1}{r|}{1}                    & \multicolumn{1}{r|}{0}                   & 16                         \\ \hline
\multicolumn{8}{|c|}{Decompression   Throughput (MB/s)}                                                                                                                                                                                                                                    \\ \hline
\multicolumn{1}{l|}{Isabel}     & 310                     & 2,601                   & \multicolumn{1}{r|}{\textbf{28,754}}  & 13                                       & \multicolumn{1}{r|}{13}                   & \multicolumn{1}{r|}{281}                 & 4                          \\
\multicolumn{1}{l|}{Tangaroa}   & 316                     & 2,592                   & \multicolumn{1}{r|}{\textbf{19,059}}  & 11                                       & \multicolumn{1}{r|}{11}                   & \multicolumn{1}{r|}{DNF}                 & 4                          \\
\multicolumn{1}{l|}{Earthquake} & 441                     & 1,896                   & \multicolumn{1}{r|}{\textbf{28,200}}  & 15                                       & \multicolumn{1}{r|}{14}                   & \multicolumn{1}{r|}{564}                 & 8                          \\
\multicolumn{1}{l|}{Ionization} & 339                     & 3,628                   & \multicolumn{1}{r|}{\textbf{33,305}}  & 13                                       & \multicolumn{1}{r|}{13}                   & \multicolumn{1}{r|}{DNF}                 & 9                          \\
\multicolumn{1}{l|}{Miranda}    & 422                     & 4,314                   & \multicolumn{1}{r|}{\textbf{30,199}}  & 10                                       & \multicolumn{1}{r|}{TO}                   & \multicolumn{1}{r|}{702}                 & 8                          \\
\multicolumn{1}{l|}{S3D}        & TO                      & 3,448                   & \multicolumn{1}{r|}{\textbf{123,457}} & 14                                       & \multicolumn{1}{r|}{TO}                   & \multicolumn{1}{r|}{446}                 & 5                          \\
\multicolumn{1}{l|}{SCALE}      & TO                      & TO                      & \multicolumn{1}{r|}{\textbf{112,896}} & 14                                       & \multicolumn{1}{r|}{TO}                   & \multicolumn{1}{r|}{649}                 & 5                          \\
\multicolumn{1}{l|}{QMCPACK}    & 429                     & 4,380                   & \multicolumn{1}{r|}{\textbf{10,950}}  & 13                                       & \multicolumn{1}{r|}{13}                   & \multicolumn{1}{r|}{438}                 & 9                          \\
\multicolumn{1}{l|}{Geomean}    & 372                     & 3,142                   & \multicolumn{1}{r|}{\textbf{35,229}}  & 13                                       & \multicolumn{1}{r|}{13}                   & \multicolumn{1}{r|}{492}                 & 6                         
\end{tabular}
}
    \end{center}
\end{table}

\begin{table}[hbtp]
    \begin{center}
        \caption{Comparison with topology-preserving compressors for the 1E-4 NOA error bound}
        \vspace{-1mm}
        \label{tab:1e-4res}
        \resizebox{1\columnwidth}{!}{
        \begin{tabular}{lrrrrrrr}
\multicolumn{1}{l|}{}           & \multicolumn{3}{c|}{LOPC}                                                                 & \multicolumn{2}{c|}{TopoA SZ3}                                                       & \multicolumn{1}{c|}{}       & \multicolumn{1}{c}{}       \\
\multicolumn{1}{l|}{}           & \multicolumn{1}{c}{Ser} & \multicolumn{1}{c}{OMP} & \multicolumn{1}{c|}{CUDA}             & \multicolumn{1}{c}{$\epsilon$ = 1.5x EB} & \multicolumn{1}{c|}{$\epsilon$ = 0.5x EB} & \multicolumn{1}{c|}{TopoSZ} & \multicolumn{1}{c}{TopoQZ} \\ \hline
\multicolumn{8}{|c|}{Compression Ratio}                                                                                                                                                                                                                                       \\ \hline
\multicolumn{1}{l|}{Isabel}     & 4.60                    & 4.60                    & \multicolumn{1}{r|}{4.60}             & \textbf{11.92}                           & \multicolumn{1}{r|}{11.51}                & \multicolumn{1}{r|}{TO}     & 2.65                       \\
\multicolumn{1}{l|}{Tangaroa}   & 4.13                    & 4.13                    & \multicolumn{1}{r|}{4.13}             & \textbf{15.49}                           & \multicolumn{1}{r|}{15.33}                & \multicolumn{1}{r|}{10.52}  & 2.35                       \\
\multicolumn{1}{l|}{Earthquake} & 7.73                    & 7.73                    & \multicolumn{1}{r|}{7.73}             & \textbf{14.51}                           & \multicolumn{1}{r|}{TO}                   & \multicolumn{1}{r|}{6.82}   & 4.85                       \\
\multicolumn{1}{l|}{Ionization} & 8.36                    & 8.36                    & \multicolumn{1}{r|}{8.36}             & \textbf{33.31}                           & \multicolumn{1}{r|}{30.78}                & \multicolumn{1}{r|}{13.78}  & 4.46                       \\
\multicolumn{1}{l|}{Miranda}    & 8.81                    & 8.81                    & \multicolumn{1}{r|}{8.81}             & \textbf{47.85}                           & \multicolumn{1}{r|}{TO}                   & \multicolumn{1}{r|}{24.63}  & 4.56                       \\
\multicolumn{1}{l|}{S3D}        & 9.11                    & 9.11                    & \multicolumn{1}{r|}{9.11}             & \textbf{51.49}                           & \multicolumn{1}{r|}{TO}                   & \multicolumn{1}{r|}{TO}     & 4.56                       \\
\multicolumn{1}{l|}{SCALE}      & \textbf{9.53}           & \textbf{9.53}           & \multicolumn{1}{r|}{\textbf{9.53}}    & 2.53                                     & \multicolumn{1}{r|}{TO}                   & \multicolumn{1}{r|}{TO}     & 3.98                       \\
\multicolumn{1}{l|}{QMCPACK}    & 8.23                    & 8.23                    & \multicolumn{1}{r|}{8.23}             & \textbf{39.60}                           & \multicolumn{1}{r|}{39.59}                & \multicolumn{1}{r|}{11.07}  & 4.66                       \\
\multicolumn{1}{l|}{Geomean}    & 7.26                    & 7.26                    & \multicolumn{1}{r|}{7.26}             & 19.63                                    & \multicolumn{1}{r|}{\textbf{21.53}}       & \multicolumn{1}{r|}{12.19}  & 3.89                       \\ \hline
\multicolumn{8}{|c|}{Compression Throughput   (MB/s)}                                                                                                                                                                                                                         \\ \hline
\multicolumn{1}{l|}{Isabel}     & 79                      & 8,182                   & \multicolumn{1}{r|}{\textbf{8,411}}   & 1                                        & \multicolumn{1}{r|}{1}                    & \multicolumn{1}{r|}{TO}     & 12                         \\
\multicolumn{1}{l|}{Tangaroa}   & 46                      & 199                     & \multicolumn{1}{r|}{\textbf{2,592}}   & 1                                        & \multicolumn{1}{r|}{1}                    & \multicolumn{1}{r|}{0}      & 13                         \\
\multicolumn{1}{l|}{Earthquake} & 83                      & 470                     & \multicolumn{1}{r|}{\textbf{7,050}}   & 2                                        & \multicolumn{1}{r|}{TO}                   & \multicolumn{1}{r|}{0}      & 21                         \\
\multicolumn{1}{l|}{Ionization} & 48                      & 102                     & \multicolumn{1}{r|}{\textbf{4,063}}   & 2                                        & \multicolumn{1}{r|}{1}                    & \multicolumn{1}{r|}{0}      & 11                         \\
\multicolumn{1}{l|}{Miranda}    & 7                       & 13                      & \multicolumn{1}{r|}{\textbf{643}}     & 1                                        & \multicolumn{1}{r|}{TO}                   & \multicolumn{1}{r|}{1}      & 8                          \\
\multicolumn{1}{l|}{S3D}        & 8                       & 568                     & \multicolumn{1}{r|}{\textbf{21,044}}  & 3                                        & \multicolumn{1}{r|}{TO}                   & \multicolumn{1}{r|}{TO}     & 22                         \\
\multicolumn{1}{l|}{SCALE}      & TO                      & 3                       & \multicolumn{1}{r|}{\textbf{179}}     & 1                                        & \multicolumn{1}{r|}{TO}                   & \multicolumn{1}{r|}{TO}     & 31                         \\
\multicolumn{1}{l|}{QMCPACK}    & 146                     & 438                     & \multicolumn{1}{r|}{\textbf{4,380}}   & 2                                        & \multicolumn{1}{r|}{2}                    & \multicolumn{1}{r|}{1}      & 13                         \\
\multicolumn{1}{l|}{Geomean}    & 38                      & 173                     & \multicolumn{1}{r|}{\textbf{3,003}}   & 1                                        & \multicolumn{1}{r|}{1}                    & \multicolumn{1}{r|}{0}      & 15                         \\ \hline
\multicolumn{8}{|c|}{Decompression   Throughput (MB/s)}                                                                                                                                                                                                                       \\ \hline
\multicolumn{1}{l|}{Isabel}     & 307                     & 2,894                   & \multicolumn{1}{r|}{\textbf{31,142}}  & 13                                       & \multicolumn{1}{r|}{13}                   & \multicolumn{1}{r|}{TO}     & 2                          \\
\multicolumn{1}{l|}{Tangaroa}   & 324                     & 2,057                   & \multicolumn{1}{r|}{\textbf{27,284}}  & 11                                       & \multicolumn{1}{r|}{11}                   & \multicolumn{1}{r|}{259}    & 4                          \\
\multicolumn{1}{l|}{Earthquake} & 470                     & 1,986                   & \multicolumn{1}{r|}{\textbf{21,793}}  & 15                                       & \multicolumn{1}{r|}{TO}                   & \multicolumn{1}{r|}{403}    & 6                          \\
\multicolumn{1}{l|}{Ionization} & 406                     & 4,515                   & \multicolumn{1}{r|}{\textbf{29,921}}  & 12                                       & \multicolumn{1}{r|}{12}                   & \multicolumn{1}{r|}{508}    & 8                          \\
\multicolumn{1}{l|}{Miranda}    & 425                     & 3,355                   & \multicolumn{1}{r|}{\textbf{33,780}}  & 10                                       & \multicolumn{1}{r|}{TO}                   & \multicolumn{1}{r|}{559}    & 7                          \\
\multicolumn{1}{l|}{S3D}        & 383                     & 3,448                   & \multicolumn{1}{r|}{\textbf{50,618}}  & 15                                       & \multicolumn{1}{r|}{TO}                   & \multicolumn{1}{r|}{TO}     & 4                          \\
\multicolumn{1}{l|}{SCALE}      & TO                      & 3,421                   & \multicolumn{1}{r|}{\textbf{112,896}} & 8                                        & \multicolumn{1}{r|}{TO}                   & \multicolumn{1}{r|}{TO}     & 3                          \\
\multicolumn{1}{l|}{QMCPACK}    & 438                     & 4,380                   & \multicolumn{1}{r|}{\textbf{10,950}}  & 13                                       & \multicolumn{1}{r|}{12}                   & \multicolumn{1}{r|}{438}    & 8                          \\
\multicolumn{1}{l|}{Geomean}    & 389                     & 3,132                   & \multicolumn{1}{r|}{\textbf{32,253}}  & 12                                       & \multicolumn{1}{r|}{12}                   & \multicolumn{1}{r|}{419}    & 5                         
\end{tabular}
}
    \end{center}
\end{table}

\begin{table*}[hbtp]
    \begin{center}
        \caption{Comparison with non-topology-preserving compressors for the 1E-2 NOA error bound}
        \vspace{-1mm}
        \label{tab:1e-2nontopres}
        \resizebox{1.33\columnwidth}{!}{
        \begin{tabular}{lrrrrrrrrr}
\multicolumn{1}{l|}{}           & \multicolumn{3}{c|}{LOPC}                                                        & \multicolumn{1}{c|}{}                  & \multicolumn{1}{c|}{}                 & \multicolumn{1}{l|}{}      & \multicolumn{1}{l|}{}     & \multicolumn{2}{c}{FPCompress}                        \\
\multicolumn{1}{l|}{}           & \multicolumn{1}{c}{Ser} & \multicolumn{1}{c}{OMP} & \multicolumn{1}{c|}{CUDA}    & \multicolumn{1}{c|}{SZ3}               & \multicolumn{1}{c|}{PFPL}             & \multicolumn{1}{c|}{FPZip} & \multicolumn{1}{c|}{ZSTD} & \multicolumn{1}{c}{Speed} & \multicolumn{1}{c}{Ratio} \\ \hline
\multicolumn{10}{|c|}{Compression Ratio}                                                                                                                                                                                                                                                                             \\ \hline
\multicolumn{1}{l|}{Isabel}     & 5.47                    & 5.47                    & \multicolumn{1}{r|}{5.47}    & \multicolumn{1}{r|}{\textbf{1,258.13}} & \multicolumn{1}{r|}{33.56}            & \multicolumn{1}{r|}{2.42}  & \multicolumn{1}{r|}{1.16} & 1.52                      & 1.73                      \\
\multicolumn{1}{l|}{Tangaroa}   & 4.39                    & 4.39                    & \multicolumn{1}{r|}{4.39}    & \multicolumn{1}{r|}{\textbf{285.80}}   & \multicolumn{1}{r|}{34.55}            & \multicolumn{1}{r|}{3.85}  & \multicolumn{1}{r|}{1.43} & 1.76                      & 2.12                      \\
\multicolumn{1}{l|}{Earthquake} & 8.79                    & 8.79                    & \multicolumn{1}{r|}{8.79}    & \multicolumn{1}{r|}{\textbf{1,336.56}} & \multicolumn{1}{r|}{102.00}           & \multicolumn{1}{r|}{1.31}  & \multicolumn{1}{r|}{1.18} & 1.19                      & 1.19                      \\
\multicolumn{1}{l|}{Ionization} & 10.19                   & 10.19                   & \multicolumn{1}{r|}{10.19}   & \multicolumn{1}{r|}{\textbf{422.57}}   & \multicolumn{1}{r|}{66.92}            & \multicolumn{1}{r|}{2.26}  & \multicolumn{1}{r|}{9.50} & 1.47                      & 2.89                      \\
\multicolumn{1}{l|}{Miranda}    & 10.05                   & 10.05                   & \multicolumn{1}{r|}{10.05}   & \multicolumn{1}{r|}{\textbf{1,350.33}} & \multicolumn{1}{r|}{94.05}            & \multicolumn{1}{r|}{2.06}  & \multicolumn{1}{r|}{2.07} & 1.84                      & 1.88                      \\
\multicolumn{1}{l|}{S3D}        & 9.48                    & 9.48                    & \multicolumn{1}{r|}{9.48}    & \multicolumn{1}{r|}{\textbf{3,097.37}} & \multicolumn{1}{r|}{63.69}            & \multicolumn{1}{r|}{1.68}  & \multicolumn{1}{r|}{1.09} & 1.33                      & 1.33                      \\
\multicolumn{1}{l|}{SCALE}      & 10.74                   & 10.74                   & \multicolumn{1}{r|}{10.74}   & \multicolumn{1}{r|}{\textbf{1,334.26}} & \multicolumn{1}{r|}{120.57}           & \multicolumn{1}{r|}{1.48}  & \multicolumn{1}{r|}{2.76} & 1.35                      & 1.36                      \\
\multicolumn{1}{l|}{QMCPACK}    & 9.07                    & 9.07                    & \multicolumn{1}{r|}{9.07}    & \multicolumn{1}{r|}{\textbf{853.99}}   & \multicolumn{1}{r|}{57.97}            & \multicolumn{1}{r|}{1.53}  & \multicolumn{1}{r|}{1.61} & 1.33                      & 1.43                      \\
\multicolumn{1}{l|}{Geomean}    & 8.18                    & 8.18                    & \multicolumn{1}{r|}{8.18}    & \multicolumn{1}{r|}{\textbf{995.92}}   & \multicolumn{1}{r|}{65.32}            & \multicolumn{1}{r|}{1.96}  & \multicolumn{1}{r|}{1.92} & 1.46                      & 1.67                      \\ \hline
\multicolumn{10}{|c|}{Compression Throughput (MB/s)}                                                                                                                                                                                                                                                                 \\ \hline
\multicolumn{1}{l|}{Isabel}     & 9                       & 17                      & \multicolumn{1}{r|}{750}     & \multicolumn{1}{r|}{268}               & \multicolumn{1}{r|}{229,479}          & \multicolumn{1}{r|}{96}    & \multicolumn{1}{r|}{4}    & \textbf{234,375}          & 194,805                   \\
\multicolumn{1}{l|}{Tangaroa}   & 1                       & 1                       & \multicolumn{1}{r|}{104}     & \multicolumn{1}{r|}{240}               & \multicolumn{1}{r|}{\textbf{207,480}} & \multicolumn{1}{r|}{130}   & \multicolumn{1}{r|}{3}    & 75,349                    & 63,066                    \\
\multicolumn{1}{l|}{Earthquake} & 7                       & 13                      & \multicolumn{1}{r|}{940}     & \multicolumn{1}{r|}{424}               & \multicolumn{1}{r|}{\textbf{166,903}} & \multicolumn{1}{r|}{130}   & \multicolumn{1}{r|}{5}    & 76,011                    & 11,515                    \\
\multicolumn{1}{l|}{Ionization} & 3                       & 4                       & \multicolumn{1}{r|}{313}     & \multicolumn{1}{r|}{469}               & \multicolumn{1}{r|}{\textbf{171,798}} & \multicolumn{1}{r|}{183}   & \multicolumn{1}{r|}{1}    & 103,390                   & 12,209                    \\
\multicolumn{1}{l|}{Miranda}    & 5                       & 8                       & \multicolumn{1}{r|}{414}     & \multicolumn{1}{r|}{460}               & \multicolumn{1}{r|}{223,927}          & \multicolumn{1}{r|}{169}   & \multicolumn{1}{r|}{2}    & \textbf{434,518}          & 9,798                     \\
\multicolumn{1}{l|}{S3D}        & TO                      & 9                       & \multicolumn{1}{r|}{408}     & \multicolumn{1}{r|}{456}               & \multicolumn{1}{r|}{241,723}          & \multicolumn{1}{r|}{164}   & \multicolumn{1}{r|}{2}    & \textbf{469,043}          & 9,859                     \\
\multicolumn{1}{l|}{SCALE}      & TO                      & TO                      & \multicolumn{1}{r|}{34}      & \multicolumn{1}{r|}{488}               & \multicolumn{1}{r|}{241,089}          & \multicolumn{1}{r|}{139}   & \multicolumn{1}{r|}{1}    & \textbf{473,557}          & 13,157                    \\
\multicolumn{1}{l|}{QMCPACK}    & 12                      & 16                      & \multicolumn{1}{r|}{730}     & \multicolumn{1}{r|}{327}               & \multicolumn{1}{r|}{\textbf{52,124}}  & \multicolumn{1}{r|}{137}   & \multicolumn{1}{r|}{6}    & 12,959                    & 5,566                     \\
\multicolumn{1}{l|}{Geomean}    & 5                       & 8                       & \multicolumn{1}{r|}{314}     & \multicolumn{1}{r|}{379}               & \multicolumn{1}{r|}{\textbf{176,190}} & \multicolumn{1}{r|}{141}   & \multicolumn{1}{r|}{3}    & 142,869                   & 18,234                    \\ \hline
\multicolumn{10}{|c|}{Decompression   Throughput (MB/s)}                                                                                                                                                                                                                                                             \\ \hline
\multicolumn{1}{l|}{Isabel}     & 310                     & 2,601                   & \multicolumn{1}{r|}{28,754}  & \multicolumn{1}{r|}{644}               & \multicolumn{1}{r|}{\textbf{314,809}} & \multicolumn{1}{r|}{60}    & \multicolumn{1}{r|}{526}  & 201,794                   & 143,770                   \\
\multicolumn{1}{l|}{Tangaroa}   & 316                     & 2,592                   & \multicolumn{1}{r|}{19,059}  & \multicolumn{1}{r|}{607}               & \multicolumn{1}{r|}{\textbf{278,159}} & \multicolumn{1}{r|}{68}    & \multicolumn{1}{r|}{298}  & 72,000                    & 62,760                    \\
\multicolumn{1}{l|}{Earthquake} & 441                     & 1,896                   & \multicolumn{1}{r|}{28,200}  & \multicolumn{1}{r|}{992}               & \multicolumn{1}{r|}{\textbf{340,120}} & \multicolumn{1}{r|}{63}    & \multicolumn{1}{r|}{513}  & 85,455                    & 126,457                   \\
\multicolumn{1}{l|}{Ionization} & 339                     & 3,628                   & \multicolumn{1}{r|}{33,305}  & \multicolumn{1}{r|}{1,034}             & \multicolumn{1}{r|}{\textbf{351,053}} & \multicolumn{1}{r|}{81}    & \multicolumn{1}{r|}{726}  & 117,775                   & 32,821                    \\
\multicolumn{1}{l|}{Miranda}    & 422                     & 4,314                   & \multicolumn{1}{r|}{30,199}  & \multicolumn{1}{r|}{979}               & \multicolumn{1}{r|}{\textbf{429,901}} & \multicolumn{1}{r|}{72}    & \multicolumn{1}{r|}{651}  & 378,433                   & 184,703                   \\
\multicolumn{1}{l|}{S3D}        & TO                      & 3,448                   & \multicolumn{1}{r|}{123,457} & \multicolumn{1}{r|}{979}               & \multicolumn{1}{r|}{392,667}          & \multicolumn{1}{r|}{68}    & \multicolumn{1}{r|}{792}  & \textbf{458,505}          & 210,084                   \\
\multicolumn{1}{l|}{SCALE}      & TO                      & TO                      & \multicolumn{1}{r|}{112,896} & \multicolumn{1}{r|}{1,009}             & \multicolumn{1}{r|}{430,160}          & \multicolumn{1}{r|}{66}    & \multicolumn{1}{r|}{495}  & \textbf{464,975}          & 211,455                   \\
\multicolumn{1}{l|}{QMCPACK}    & 429                     & 4,380                   & \multicolumn{1}{r|}{10,950}  & \multicolumn{1}{r|}{693}               & \multicolumn{1}{r|}{\textbf{164,518}} & \multicolumn{1}{r|}{17}    & \multicolumn{1}{r|}{17}   & 14,504                    & 54,752                    \\
\multicolumn{1}{l|}{Geomean}    & 372                     & 3,142                   & \multicolumn{1}{r|}{35,229}  & \multicolumn{1}{r|}{849}               & \multicolumn{1}{r|}{\textbf{325,142}} & \multicolumn{1}{r|}{57}    & \multicolumn{1}{r|}{354}  & 142,613                   & 106,719                  
\end{tabular}
}
    \end{center}
\end{table*}

\begin{table*}[hbtp]
    \begin{center}
        \caption{Comparison with non-topology-preserving compressors for the 1E-4 NOA error bound}
        \vspace{-1mm}
        \label{tab:1e-4nontopres}
        \resizebox{1.33\columnwidth}{!}{
        \begin{tabular}{lrrrrrrrrr}
\multicolumn{1}{l|}{}           & \multicolumn{3}{c|}{LOPC}                                                        & \multicolumn{1}{c|}{}                & \multicolumn{1}{c|}{}                 & \multicolumn{1}{l|}{}      & \multicolumn{1}{l|}{}     & \multicolumn{2}{c}{FPCompress}                        \\
\multicolumn{1}{l|}{}           & \multicolumn{1}{c}{Ser} & \multicolumn{1}{c}{OMP} & \multicolumn{1}{c|}{CUDA}    & \multicolumn{1}{c|}{SZ3}             & \multicolumn{1}{c|}{PFPL}             & \multicolumn{1}{c|}{FPZip} & \multicolumn{1}{c|}{ZSTD} & \multicolumn{1}{c}{Speed} & \multicolumn{1}{c}{Ratio} \\ \hline
\multicolumn{10}{|c|}{Compression Ratio}                                                                                                                                                                                                                                                                           \\ \hline
\multicolumn{1}{l|}{Isabel}     & 4.60                    & 4.60                    & \multicolumn{1}{r|}{4.60}    & \multicolumn{1}{r|}{\textbf{23.06}}  & \multicolumn{1}{r|}{6.90}             & \multicolumn{1}{r|}{2.42}  & \multicolumn{1}{r|}{1.16} & 1.52                      & 1.73                      \\
\multicolumn{1}{l|}{Tangaroa}   & 4.13                    & 4.13                    & \multicolumn{1}{r|}{4.13}    & \multicolumn{1}{r|}{\textbf{24.49}}  & \multicolumn{1}{r|}{7.32}             & \multicolumn{1}{r|}{3.85}  & \multicolumn{1}{r|}{1.43} & 1.76                      & 2.12                      \\
\multicolumn{1}{l|}{Earthquake} & 7.73                    & 7.73                    & \multicolumn{1}{r|}{7.73}    & \multicolumn{1}{r|}{\textbf{33.99}}  & \multicolumn{1}{r|}{13.97}            & \multicolumn{1}{r|}{1.31}  & \multicolumn{1}{r|}{1.18} & 1.19                      & 1.19                      \\
\multicolumn{1}{l|}{Ionization} & 8.36                    & 8.36                    & \multicolumn{1}{r|}{8.36}    & \multicolumn{1}{r|}{\textbf{54.59}}  & \multicolumn{1}{r|}{16.29}            & \multicolumn{1}{r|}{2.26}  & \multicolumn{1}{r|}{9.50} & 1.47                      & 2.89                      \\
\multicolumn{1}{l|}{Miranda}    & 8.81                    & 8.81                    & \multicolumn{1}{r|}{8.81}    & \multicolumn{1}{r|}{\textbf{81.92}}  & \multicolumn{1}{r|}{30.34}            & \multicolumn{1}{r|}{2.06}  & \multicolumn{1}{r|}{2.07} & 1.84                      & 1.88                      \\
\multicolumn{1}{l|}{S3D}        & 9.11                    & 9.11                    & \multicolumn{1}{r|}{9.11}    & \multicolumn{1}{r|}{\textbf{127.02}} & \multicolumn{1}{r|}{13.91}            & \multicolumn{1}{r|}{1.68}  & \multicolumn{1}{r|}{1.09} & 1.33                      & 1.33                      \\
\multicolumn{1}{l|}{SCALE}      & 9.53                    & 9.53                    & \multicolumn{1}{r|}{9.53}    & \multicolumn{1}{r|}{\textbf{30.03}}  & \multicolumn{1}{r|}{20.93}            & \multicolumn{1}{r|}{1.48}  & \multicolumn{1}{r|}{2.76} & 1.35                      & 1.36                      \\
\multicolumn{1}{l|}{QMCPACK}    & 8.23                    & 8.23                    & \multicolumn{1}{r|}{8.23}    & \multicolumn{1}{r|}{\textbf{80.89}}  & \multicolumn{1}{r|}{12.58}            & \multicolumn{1}{r|}{1.53}  & \multicolumn{1}{r|}{1.61} & 1.33                      & 1.43                      \\
\multicolumn{1}{l|}{Geomean}    & 7.26                    & 7.26                    & \multicolumn{1}{r|}{7.26}    & \multicolumn{1}{r|}{\textbf{47.63}}  & \multicolumn{1}{r|}{13.75}            & \multicolumn{1}{r|}{1.96}  & \multicolumn{1}{r|}{1.92} & 1.46                      & 1.67                      \\ \hline
\multicolumn{10}{|c|}{Compression Throughput (MB/s)}                                                                                                                                                                                                                                                               \\ \hline
\multicolumn{1}{l|}{Isabel}     & 79                      & 8,182                   & \multicolumn{1}{r|}{8,411}   & \multicolumn{1}{r|}{195}             & \multicolumn{1}{r|}{225,361}          & \multicolumn{1}{r|}{96}    & \multicolumn{1}{r|}{4}    & \textbf{234,375}          & 194,805                   \\
\multicolumn{1}{l|}{Tangaroa}   & 46                      & 199                     & \multicolumn{1}{r|}{2,592}   & \multicolumn{1}{r|}{206}             & \multicolumn{1}{r|}{\textbf{205,793}} & \multicolumn{1}{r|}{130}   & \multicolumn{1}{r|}{3}    & 75,349                    & 63,066                    \\
\multicolumn{1}{l|}{Earthquake} & 83                      & 470                     & \multicolumn{1}{r|}{7,050}   & \multicolumn{1}{r|}{329}             & \multicolumn{1}{r|}{\textbf{162,953}} & \multicolumn{1}{r|}{130}   & \multicolumn{1}{r|}{5}    & 76,011                    & 11,515                    \\
\multicolumn{1}{l|}{Ionization} & 48                      & 102                     & \multicolumn{1}{r|}{4,063}   & \multicolumn{1}{r|}{384}             & \multicolumn{1}{r|}{\textbf{170,300}} & \multicolumn{1}{r|}{183}   & \multicolumn{1}{r|}{1}    & 103,390                   & 12,209                    \\
\multicolumn{1}{l|}{Miranda}    & 7                       & 13                      & \multicolumn{1}{r|}{643}     & \multicolumn{1}{r|}{434}             & \multicolumn{1}{r|}{223,080}          & \multicolumn{1}{r|}{169}   & \multicolumn{1}{r|}{2}    & \textbf{434,518}          & 9,798                     \\
\multicolumn{1}{l|}{S3D}        & 8                       & 568                     & \multicolumn{1}{r|}{21,044}  & \multicolumn{1}{r|}{407}             & \multicolumn{1}{r|}{229,240}          & \multicolumn{1}{r|}{164}   & \multicolumn{1}{r|}{2}    & \textbf{469,043}          & 9,859                     \\
\multicolumn{1}{l|}{SCALE}      & TO                      & 3                       & \multicolumn{1}{r|}{179}     & \multicolumn{1}{r|}{376}             & \multicolumn{1}{r|}{244,565}          & \multicolumn{1}{r|}{139}   & \multicolumn{1}{r|}{1}    & \textbf{473,557}          & 13,157                    \\
\multicolumn{1}{l|}{QMCPACK}    & 146                     & 438                     & \multicolumn{1}{r|}{4,380}   & \multicolumn{1}{r|}{251}             & \multicolumn{1}{r|}{\textbf{49,738}}  & \multicolumn{1}{r|}{137}   & \multicolumn{1}{r|}{6}    & 12,959                    & 5,566                     \\
\multicolumn{1}{l|}{Geomean}    & 38                      & 173                     & \multicolumn{1}{r|}{3,003}   & \multicolumn{1}{r|}{310}             & \multicolumn{1}{r|}{\textbf{172,953}} & \multicolumn{1}{r|}{141}   & \multicolumn{1}{r|}{3}    & 142,869                   & 18,234                    \\ \hline
\multicolumn{10}{|c|}{Decompression   Throughput (MB/s)}                                                                                                                                                                                                                                                           \\ \hline
\multicolumn{1}{l|}{Isabel}     & 307                     & 2,894                   & \multicolumn{1}{r|}{31,142}  & \multicolumn{1}{r|}{416}             & \multicolumn{1}{r|}{\textbf{304,679}} & \multicolumn{1}{r|}{60}    & \multicolumn{1}{r|}{526}  & 201,794                   & 143,770                   \\
\multicolumn{1}{l|}{Tangaroa}   & 324                     & 2,057                   & \multicolumn{1}{r|}{27,284}  & \multicolumn{1}{r|}{446}             & \multicolumn{1}{r|}{\textbf{275,136}} & \multicolumn{1}{r|}{68}    & \multicolumn{1}{r|}{298}  & 72,000                    & 62,760                    \\
\multicolumn{1}{l|}{Earthquake} & 470                     & 1,986                   & \multicolumn{1}{r|}{21,793}  & \multicolumn{1}{r|}{652}             & \multicolumn{1}{r|}{\textbf{335,459}} & \multicolumn{1}{r|}{63}    & \multicolumn{1}{r|}{513}  & 85,455                    & 126,457                   \\
\multicolumn{1}{l|}{Ionization} & 406                     & 4,515                   & \multicolumn{1}{r|}{29,921}  & \multicolumn{1}{r|}{772}             & \multicolumn{1}{r|}{\textbf{348,070}} & \multicolumn{1}{r|}{81}    & \multicolumn{1}{r|}{726}  & 117,775                   & 32,821                    \\
\multicolumn{1}{l|}{Miranda}    & 425                     & 3,355                   & \multicolumn{1}{r|}{33,780}  & \multicolumn{1}{r|}{835}             & \multicolumn{1}{r|}{\textbf{426,173}} & \multicolumn{1}{r|}{72}    & \multicolumn{1}{r|}{651}  & 378,433                   & 184,703                   \\
\multicolumn{1}{l|}{S3D}        & 383                     & 3,448                   & \multicolumn{1}{r|}{50,618}  & \multicolumn{1}{r|}{864}             & \multicolumn{1}{r|}{384,781}          & \multicolumn{1}{r|}{68}    & \multicolumn{1}{r|}{792}  & \textbf{458,505}          & 210,084                   \\
\multicolumn{1}{l|}{SCALE}      & TO                      & 3,421                   & \multicolumn{1}{r|}{112,896} & \multicolumn{1}{r|}{608}             & \multicolumn{1}{r|}{402,804}          & \multicolumn{1}{r|}{66}    & \multicolumn{1}{r|}{495}  & \textbf{464,975}          & 211,455                   \\
\multicolumn{1}{l|}{QMCPACK}    & 438                     & 4,380                   & \multicolumn{1}{r|}{10,950}  & \multicolumn{1}{r|}{602}             & \multicolumn{1}{r|}{\textbf{164,518}} & \multicolumn{1}{r|}{17}    & \multicolumn{1}{r|}{17}   & 14,504                    & 54,752                    \\
\multicolumn{1}{l|}{Geomean}    & 389                     & 3,132                   & \multicolumn{1}{r|}{32,253}  & \multicolumn{1}{r|}{630}             & \multicolumn{1}{r|}{\textbf{318,677}} & \multicolumn{1}{r|}{57}    & \multicolumn{1}{r|}{354}  & 142,613                   & 106,719                  
\end{tabular}
        }
    \end{center}
\end{table*}

The compression ratios of the evaluated compressors across the two tested error bounds are shown in the first section of Tables~\ref{tab:1e-2res}, \ref{tab:1e-4res}, \ref{tab:1e-2nontopres}, and~\ref{tab:1e-4nontopres}. Due to its CPU/GPU parity guarantee, LOPC maintains the same compression ratio across both devices.

Compared to the other compressors that maintain local order information, namely the lossless compressors (FPZip, ZSTD, and FPCompress), LOPC compresses more in all but one case. On the Ionization input with an error bound of 1E-4, LOPC's compression ratio is a little lower than that of ZSTD. In all other cases, its compression ratio is higher, on average by a factor of 3.7 and, in one case, by over a factor of 8.7.

In general, LOPC yields lower compression ratios than the other topology preserving compressors and significantly lower compression ratios than the lossy non-topology preserving compressors. This is expected for the following three reasons.

First, because LOPC preserves all local ordering, it often preserves many relationships that the other topology preserving compressors do not (see Section~\ref{subsec:crit_points}). Of course, it preserves even more relationships than the lossy non-topology-preserving compressors. This large amount of extra information is the main reason for the lower compression ratio.

Second, by storing bins and subbins separately, the information density of each part of the compressed file changes with the user-requested error bound. For example, if the user requests a strict error bound, most of the topological information ends up in the quantization bins, making them more challenging to compress easily, while the subbins contain almost no information. The reverse is true for a loose error bound because the topological information must now be stored in the subbins. This makes it difficult for a single compression algorithm to be effective in all cases due to the greatly changing information density. We discuss this relationship more in Section~\ref{subsec:eb_and_perf}.

Third, LOPC is designed to work on both CPUs and GPUs. As a consequence, its compression algorithm cannot exploit some of the effective serial transformations that the CPU-only compressors include. %such as Huffman encoding~\cite{huffman}.

\subsection{Compression and Decompression Speed}

The speeds at which the evaluated compressors compress and decompress our inputs are listed in the middle and last section of Tables~\ref{tab:1e-2res}, \ref{tab:1e-4res}, \ref{tab:1e-2nontopres}, and~\ref{tab:1e-4nontopres}, respectively. The speed of the topology-preserving compressors is very input dependent. For example, even for LOPC, the serial and OpenMP versions time out on the SCALE input. Additionally, for the other topology-preserving compressors, the persistence threshold $\epsilon$ has a large impact on the runtime. TopoA augmented SZ3 with a persistence threshold less than the error bound times out on 3 of the 8 inputs for the 1E-2 error bound and on 4 of the 8 inputs for the 1E-4 error bound. These results highlight the need for fast (parallelized) topology preserving compressors like LOPC. We further discuss the effect that error bound and persistence threshold have on performance in Section~\ref{subsec:eb_and_perf}.

On the SCALE input at an error bound of 1E-4, only LOPC and TopoA with a large persistence threshold are able to finish within an hour. In this case, however, TopoA takes 22 minutes whereas LOPC only takes 6.3 seconds on the GPU, making it 209.9 times faster than TopoA. Further, TopoA is only able to compress this input using the aforementioned large persistence threshold, which causes fewer critical points to be preserved on the contour tree, while LOPC is able to preserve all critical points. Serially, LOPC performs better than the other topology preserving compressors as well. This is likely due to the other compressors building a new contour tree in each iteration, a step that LOPC is able to forgo due to its entirely different approach for preserving the topology.

Naturally, LOPC is slower than the non-topology-preserving compressors, though its CUDA version is on par with some of the serial compressors. Compared to other GPU-based compressors like PFPL and FPCompress that do not need to preserve topology, LOPC can be slower by over a factor of 100 (e.g., on the S3D input). This is expected because LOPC performs most of its topology preservation work during compression %While compression is generally performed only once and the result saved in a central location, decompression is the common case.
and explains why its decompression performance is much closer to the non-topology-preserving compressors. %LOPC yields up to 32 GB/s, on average, in terms of throughput all while preserving all local order information and critical points.

\subsection{Effects of Error Bound on LOPC Compression}
\label{subsec:eb_and_perf}

The user-requested error bound affects all tested lossy compressors in terms of both compression ratio and compression/decompression speed. In this section, we discuss how each compressor behaves when the error bound is changed. Lossy compressors that do not preserve topology tend to produce significantly lower compression ratios for smaller error bounds but only incur minor slowdowns~\cite{mb-ipdps25a}. For these compressors, the computation is not directly affected by the error bound, and the slowdown stems from the increased amount of data that needs to be written due to the lower compressibility with tighter error bounds.

\begin{figure}[!htbp]
\centering
\begin{tikzpicture}
\pgfplotsset{
    legend style={at={(.98,.35)},anchor=south east},
}
% CR
\begin{axis}[
    axis y line*=left,
    ybar,
    bar width=15pt,
	nodes near coords,
    bar width=15pt,
    ymin=1.5, ymax=12, % make start at 0, move legend into chart, turn of CR gridlines
    enlargelimits=0.15,
    ylabel={Compression Ratio},
    symbolic x coords={1E-0, 1E-1, 1E-2, 1E-3, 1E-4, 1E-5, 1E-6},
    xtick=data,
    x tick label style={rotate=45,anchor=east},
    ]
\addplot[ybar, ybar legend, fill=blue!20] coordinates {
    (1E-0, 9.34)
    (1E-1, 9.58)
    (1E-2, 10.40)
    (1E-3, 10.77)
    (1E-4, 8.78)
    (1E-5, 7.01)
    (1E-6, 5.58)
}; \label{crplot}
\end{axis}

% Runtime
\begin{axis}[
  grid=major,
  axis y line*=right,
  axis x line=none,
  ymin=-0.1, ymax=0.5, % make this look good too
  ylabel= Runtime (s),
  enlargelimits=0.15,
  legend style={at={(0.5,-0.15)},
      anchor=north,legend columns=-1},
  symbolic x coords={1E-0, 1E-1, 1E-2, 1E-3, 1E-4, 1E-5, 1E-6},
]
\addlegendimage{/pgfplots/refstyle=crplot}\addlegendentry{Compression ratio}
\addplot[thick, red]
  coordinates{
    (1E-0, 0.417)
    (1E-1, 0.271)
    (1E-2, 0.106)
    (1E-3, 0.056)
    (1E-4, 0.032)
    (1E-5, 0.021)
    (1E-6, 0.016)
}; \addlegendentry{Runtime}
\end{axis}
\end{tikzpicture}
\vspace{-1mm}
\caption{Geometric-mean compression ratio and compression runtime of LOPC on 7 NOA error bounds.}
\label{fig:cr-comp-speed-multi-eb}
\end{figure}
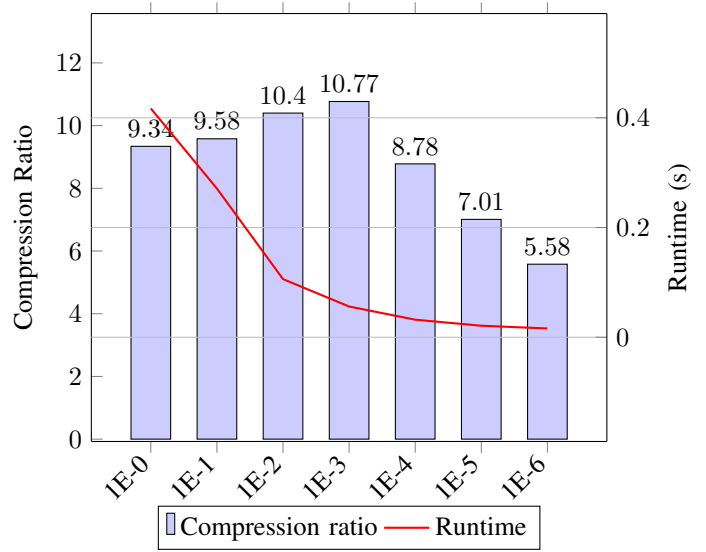

For the topology-preserving compressors, the compression speed is largely determined by how much correction needs to be performed. Figure~\ref{fig:cr-comp-speed-multi-eb} illustrates this for LOPC on 7 error bounds. For the loosest tested error bound, where almost all information is lost, LOPC must perform a large amount of correction and, thus, has the highest runtime. In contrast, the tightest tested error bound yields the lowest runtime because LOPC is ``correcting'' data that already has many of the local-order relationships intact.

 % LOPC speed and CR over the error bounds

Similar runtime behavior is observed in Tables~\ref{tab:1e-2res}~and~\ref{tab:1e-4res} for the compressors that preserve the contour tree within a persistence threshold. These compressors have an additional parameter for the persistence threshold, which also affects how much correction must be performed. %The persistence threshold has a similar effect to the error bound in that 
In particular, more correction must be done and, therefore, more time is taken when the persistence threshold is smaller than the error bound.

%In addition to the runtime, 
Figure~\ref{fig:cr-comp-speed-multi-eb} includes LOPC's compression ratio. The relationship between the error bound and the compression ratio is not as straightforward as the runtime relationship. The geometric-mean compression ratio is $9.3$ at an error bound of 1, increases to a maximum of $10.8$ at an error bound of 1E-3, and then decreases to $5.6$ at the lowest error bound of 1E-6. This behavior is due to the dual-compression design of LOPC. Since LOPC processes the bins and subbins separately, there is an optimal error bound where the information is most evenly split between the bin and subbin values. TopoA behaves similarly, where the best compressing base compressor does not always yield the highest final compression ratio~\cite{gorski2025general}.

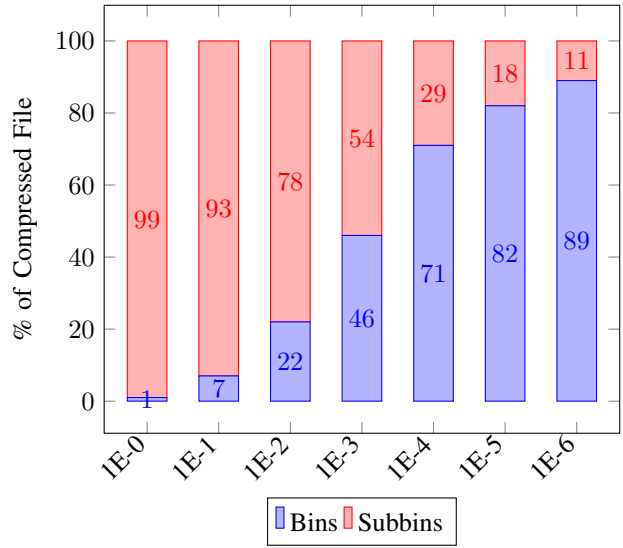
\begin{figure}[!htbp]
\centering
\begin{tikzpicture}
\begin{axis}[
    ybar stacked,
	bar width=15pt,
	nodes near coords,
    enlargelimits=0.10,
    legend style={at={(0.5,-0.15)},
      anchor=north,legend columns=-1},
    ylabel={\% of Compressed File},
    symbolic x coords={1E-0, 1E-1, 1E-2, 1E-3, 1E-4, 1E-5, 1E-6},
    xtick=data,
    x tick label style={rotate=45,anchor=east},
    ]
    % Bins
\addplot+[ybar, ybar legend] plot coordinates {
    (1E-0, 1)
    (1E-1, 7)
    (1E-2, 22)
    (1E-3, 46)
    (1E-4, 71)
    (1E-5, 82)
    (1E-6, 89)
};
    % Subbins
\addplot+[ybar, ybar legend] plot coordinates {
    (1E-0, 99)
    (1E-1, 93)
    (1E-2, 78)
    (1E-3, 54)
    (1E-4, 29)
    (1E-5, 18)
    (1E-6, 11)
};
\legend{\strut Bins, \strut Subbins}
\end{axis}
\end{tikzpicture}
\vspace{-1mm}
\caption{Average portion of the compressed file that is bin data and subbin data for LOPC on 7 NOA error bounds.}
\label{fig:cr-bins-subbins}
\end{figure} % LOPC CR breakdown over the error bounds

Figure~\ref{fig:cr-bins-subbins} shows the average fraction of the compressed file taken up by the compressed bin data and the compressed subbin data across the 7 tested error bounds. These results help explain the compression ratio results from Figure~\ref{fig:cr-comp-speed-multi-eb}. As discussed, at the loosest error bound of 1, the main data (i.e., the bin information) is almost entirely lost. Hence, the bin data is extremely compressible as it contains almost no information. In contrast, the subbins contain the corrections for the main data to maintain local order. For this reason, the subbins are much less compressible and make up 99\% of the compressed file. However, the overall compression ratio is still high because the subbins only contain small numbers that are relatively easy to compress. % Want to say something about how this means that the data required for local order is less complex than just using a tight error bound, but not really sure how to word it. Something about how we control how the subbins are assigned and encoded so we made them such that they are easier to compress vs being at the whim of the random mantissas of a float?

As the error bound decreases, the bin data starts to take up a larger portion of the compressed file. The highest observed compression ratio is seen at an error bound of 1E-3, where the bins and subbins take up an approximately equal portion of the compressed file. As discussed in Section~\ref{sec:appr}, we used the LC tool to generate good the compression algorithms for the bins and subbins. We targeted this search around the commonly used 1E-3 error bound, which contributes to the compression ratio peaking for LOPC at 1E-3.

The worst compression ratios are seen at the tightest error bounds, where the bins take up most of the compressed file. This is because at these error bounds, most of the information is stored in the bins. In fact, we are approaching lossless compression with these tight error bounds, meaning most of the original floating-point information is actually retained.
%What is more, this is closer to the original floating-point data than the subbins which greatly hinders compressibility.

In summary, all lossy compressors are affected by changes in the user-provided parameters. The non-topology-preserving lossy compressors, which tend to be memory bound, are mainly affected in compression ratio, with only a slight slowdown at lower error bounds. In contrast, the compute-bound topology-preserving compressors actually run faster at tighter error bounds, even while producing lower compression ratios, due to the decreased amount of topology correction that must be performed.

\subsection{Quality of Reconstructed Data}

Tables~\ref{tab:1e-2psnr} and~\ref{tab:1e-4psnr} show the quality of the reconstructed data yielded by each tested compressor. We show peak signal-to-noise ratio (PSNR) and the structural similarity index measure (SSIM). Both are higher-is-better quality metrics that are commonly used when evaluating reconstructed lossy data. %Table~\ref{tab:1e-4psnr} does not list MSE, as it was `0' for all tested compressors at this error bound.

Overall, TopoA augmented SZ3 produces the highest-quality reconstructed data, followed by LOPC. This is explained by the differences in how TopoA fixes errors compared to LOPC. When TopoA finds a topology problem, it tightens the error bound, bringing that value closer to the original, which necessarily moves the values closer to lossless encoding. This is in contrast to LOPC, which fixes problems with local ordering by monotonically increasing subbin values. This means that, while the local order is preserved, the individual values may be shifted further from their original value than traditional quantization. Even with this consideration, LOPC consistently produces higher quality reconstructions than the other topology-preserving compressors all while exceeding their speed by a large margin.

% 1E-2
\begin{table*}[hbtp]
    \begin{center}
        \caption{Comparison of PSNR and SSIM for the 1E-2 NOA error bound}
        \vspace{-1mm}
        \label{tab:1e-2psnr}
        \resizebox{1.31\columnwidth}{!}{
\begin{tabular}{lrrrrrrr}
\multicolumn{1}{l|}{}           & \multicolumn{1}{c|}{}              & \multicolumn{2}{c|}{TopoA SZ3}                                                       & \multicolumn{1}{c|}{}              & \multicolumn{1}{c|}{}              & \multicolumn{1}{c|}{}              & \multicolumn{1}{c}{}     \\
\multicolumn{1}{l|}{}           & \multicolumn{1}{c|}{LOPC}          & \multicolumn{1}{c}{$\epsilon$ = 1.5x EB} & \multicolumn{1}{c|}{$\epsilon$ = 0.5x EB} & \multicolumn{1}{c|}{TopoSZ}        & \multicolumn{1}{c|}{TopoQZ}        & \multicolumn{1}{c|}{SZ3}           & \multicolumn{1}{c}{PFPL} \\ \hline
\multicolumn{8}{|c|}{PSNR}                                                                                                                                                                                                                                                                            \\ \hline
\multicolumn{1}{l|}{Isabel}     & \multicolumn{1}{r|}{52.9}          & 55.2                                     & \multicolumn{1}{r|}{\textbf{57.3}}        & \multicolumn{1}{r|}{50.4}          & \multicolumn{1}{r|}{53.0}          & \multicolumn{1}{r|}{50.8}          & 44.8                     \\
\multicolumn{1}{l|}{Tangaroa}   & \multicolumn{1}{r|}{53.0}          & 56.7                                     & \multicolumn{1}{r|}{\textbf{58.4}}        & \multicolumn{1}{r|}{DNF}           & \multicolumn{1}{r|}{57.7}          & \multicolumn{1}{r|}{53.0}          & 45.0                     \\
\multicolumn{1}{l|}{Earthquake} & \multicolumn{1}{r|}{53.3}          & 59.3                                     & \multicolumn{1}{r|}{\textbf{63.0}}        & \multicolumn{1}{r|}{47.8}          & \multicolumn{1}{r|}{50.0}          & \multicolumn{1}{r|}{53.1}          & 47.4                     \\
\multicolumn{1}{l|}{Ionization} & \multicolumn{1}{r|}{51.4}          & 54.7                                     & \multicolumn{1}{r|}{\textbf{54.7}}        & \multicolumn{1}{r|}{DNF}           & \multicolumn{1}{r|}{50.0}          & \multicolumn{1}{r|}{53.9}          & 49.2                     \\
\multicolumn{1}{l|}{Miranda}    & \multicolumn{1}{r|}{\textbf{59.9}} & 46.3                                     & \multicolumn{1}{r|}{TO}                   & \multicolumn{1}{r|}{41.1}          & \multicolumn{1}{r|}{49.1}          & \multicolumn{1}{r|}{54.7}          & 52.0                     \\
\multicolumn{1}{l|}{S3D}        & \multicolumn{1}{r|}{56.0}          & 62.6                                     & \multicolumn{1}{r|}{TO}                   & \multicolumn{1}{r|}{11.7}          & \multicolumn{1}{r|}{\textbf{64.4}} & \multicolumn{1}{r|}{54.2}          & 44.7                     \\
\multicolumn{1}{l|}{SCALE}      & \multicolumn{1}{r|}{\textbf{56.3}} & 49.1                                     & \multicolumn{1}{r|}{TO}                   & \multicolumn{1}{r|}{8.0}           & \multicolumn{1}{r|}{51.2}          & \multicolumn{1}{r|}{51.2}          & 46.5                     \\
\multicolumn{1}{l|}{QMCPACK}    & \multicolumn{1}{r|}{53.1}          & 55.3                                     & \multicolumn{1}{r|}{\textbf{60.6}}        & \multicolumn{1}{r|}{49.6}          & \multicolumn{1}{r|}{51.5}          & \multicolumn{1}{r|}{47.6}          & 44.5                     \\
\multicolumn{1}{l|}{Geomean}    & \multicolumn{1}{r|}{54.4}          & 54.7                                     & \multicolumn{1}{r|}{\textbf{58.7}}        & \multicolumn{1}{r|}{27.8}          & \multicolumn{1}{r|}{53.1}          & \multicolumn{1}{r|}{52.3}          & 46.7                     \\ \hline
\multicolumn{8}{|c|}{SSIM}                                                                                                                                                                                                                                                                            \\ \hline
\multicolumn{1}{l|}{Isabel}     & \multicolumn{1}{r|}{\textbf{1.00}} & \textbf{1.00}                            & \multicolumn{1}{r|}{\textbf{1.00}}        & \multicolumn{1}{r|}{\textbf{1.00}} & \multicolumn{1}{r|}{\textbf{1.00}} & \multicolumn{1}{r|}{\textbf{1.00}} & 0.97                     \\
\multicolumn{1}{l|}{Tangaroa}   & \multicolumn{1}{r|}{\textbf{1.00}} & \textbf{1.00}                            & \multicolumn{1}{r|}{\textbf{1.00}}        & \multicolumn{1}{r|}{DNF}           & \multicolumn{1}{r|}{\textbf{1.00}} & \multicolumn{1}{r|}{\textbf{1.00}} & 0.97                     \\
\multicolumn{1}{l|}{Earthquake} & \multicolumn{1}{r|}{0.92}          & 0.98                                     & \multicolumn{1}{r|}{\textbf{0.99}}        & \multicolumn{1}{r|}{0.75}          & \multicolumn{1}{r|}{0.82}          & \multicolumn{1}{r|}{0.94}          & 0.71                     \\
\multicolumn{1}{l|}{Ionization} & \multicolumn{1}{r|}{\textbf{1.00}} & \textbf{1.00}                            & \multicolumn{1}{r|}{\textbf{1.00}}        & \multicolumn{1}{r|}{DNF}           & \multicolumn{1}{r|}{\textbf{1.00}} & \multicolumn{1}{r|}{\textbf{1.00}} & 1.00                     \\
\multicolumn{1}{l|}{Miranda}    & \multicolumn{1}{r|}{\textbf{0.31}} & 0.26                                     & \multicolumn{1}{r|}{TO}                   & \multicolumn{1}{r|}{0.29}          & \multicolumn{1}{r|}{0.30}          & \multicolumn{1}{r|}{0.27}          & 0.31                     \\
\multicolumn{1}{l|}{S3D}        & \multicolumn{1}{r|}{0.69}          & 0.95                                     & \multicolumn{1}{r|}{TO}                   & \multicolumn{1}{r|}{0.01}          & \multicolumn{1}{r|}{0.94}          & \multicolumn{1}{r|}{\textbf{0.98}} & 0.44                     \\
\multicolumn{1}{l|}{SCALE}      & \multicolumn{1}{r|}{0.62}          & \textbf{0.68}                            & \multicolumn{1}{r|}{TO}                   & \multicolumn{1}{r|}{0.00}          & \multicolumn{1}{r|}{0.62}          & \multicolumn{1}{r|}{0.67}          & 0.59                     \\
\multicolumn{1}{l|}{QMCPACK}    & \multicolumn{1}{r|}{0.99}          & 1.00                                     & \multicolumn{1}{r|}{\textbf{1.00}}        & \multicolumn{1}{r|}{0.98}          & \multicolumn{1}{r|}{0.99}          & \multicolumn{1}{r|}{0.97}          & 0.95                     \\
\multicolumn{1}{l|}{Geomean}    & \multicolumn{1}{r|}{0.77}          & 0.80                                     & \multicolumn{1}{r|}{\textbf{1.00}}        & \multicolumn{1}{r|}{0.14}          & \multicolumn{1}{r|}{0.78}          & \multicolumn{1}{r|}{0.80}          & 0.69                    
\end{tabular}
}
\end{center}
\end{table*}

% 1E-4
\begin{table*}[hbtp]
    \begin{center}
        \caption{Comparison of PSNR and SSIM for the 1E-4 NOA error bound}
        \vspace{-1mm}
        \label{tab:1e-4psnr}
        \resizebox{1.33\columnwidth}{!}{
\begin{tabular}{lrrrrrrr}
\multicolumn{1}{l|}{}           & \multicolumn{1}{c|}{}              & \multicolumn{2}{c|}{TopoA SZ3}                                                       & \multicolumn{1}{c|}{}              & \multicolumn{1}{c|}{}              & \multicolumn{1}{c|}{}              & \multicolumn{1}{c}{}     \\
\multicolumn{1}{l|}{}           & \multicolumn{1}{c|}{LOPC}          & \multicolumn{1}{c}{$\epsilon$ = 1.5x EB} & \multicolumn{1}{c|}{$\epsilon$ = 0.5x EB} & \multicolumn{1}{c|}{TopoSZ}        & \multicolumn{1}{c|}{TopoQZ}        & \multicolumn{1}{c|}{SZ3}           & \multicolumn{1}{c}{PFPL} \\ \hline
\multicolumn{8}{|c|}{PSNR}                                                                                                                                                                                                                                                                            \\ \hline
\multicolumn{1}{l|}{Isabel}     & \multicolumn{1}{r|}{\textbf{95.1}} & 94.6                                     & \multicolumn{1}{r|}{94.4}                 & \multicolumn{1}{r|}{TO}            & \multicolumn{1}{r|}{91.5}          & \multicolumn{1}{r|}{86.4}          & 84.8                     \\
\multicolumn{1}{l|}{Tangaroa}   & \multicolumn{1}{r|}{95.1}          & \textbf{96.7}                            & \multicolumn{1}{r|}{96.6}                 & \multicolumn{1}{r|}{90.4}          & \multicolumn{1}{r|}{90.9}          & \multicolumn{1}{r|}{87.8}          & 84.8                     \\
\multicolumn{1}{l|}{Earthquake} & \multicolumn{1}{r|}{\textbf{95.1}} & 94.5                                     & \multicolumn{1}{r|}{TO}                   & \multicolumn{1}{r|}{90.4}          & \multicolumn{1}{r|}{91.8}          & \multicolumn{1}{r|}{85.4}          & 84.7                     \\
\multicolumn{1}{l|}{Ionization} & \multicolumn{1}{r|}{95.7}          & 97.0                                     & \multicolumn{1}{r|}{\textbf{97.8}}        & \multicolumn{1}{r|}{88.7}          & \multicolumn{1}{r|}{90.5}          & \multicolumn{1}{r|}{89.4}          & 86.4                     \\
\multicolumn{1}{l|}{Miranda}    & \multicolumn{1}{r|}{\textbf{95.6}} & 94.0                                     & \multicolumn{1}{r|}{TO}                   & \multicolumn{1}{r|}{83.2}          & \multicolumn{1}{r|}{89.3}          & \multicolumn{1}{r|}{89.6}          & 90.0                     \\
\multicolumn{1}{l|}{S3D}        & \multicolumn{1}{r|}{92.2}          & \textbf{93.4}                            & \multicolumn{1}{r|}{TO}                   & \multicolumn{1}{r|}{TO}            & \multicolumn{1}{r|}{91.2}          & \multicolumn{1}{r|}{89.4}          & 84.8                     \\
\multicolumn{1}{l|}{SCALE}      & \multicolumn{1}{r|}{91.1}          & inf                                      & \multicolumn{1}{r|}{TO}                   & \multicolumn{1}{r|}{TO}            & \multicolumn{1}{r|}{\textbf{96.0}} & \multicolumn{1}{r|}{86.7}          & 84.4                     \\
\multicolumn{1}{l|}{QMCPACK}    & \multicolumn{1}{r|}{95.1}          & 99.3                                     & \multicolumn{1}{r|}{\textbf{99.3}}        & \multicolumn{1}{r|}{90.8}          & \multicolumn{1}{r|}{90.9}          & \multicolumn{1}{r|}{88.9}          & 84.8                     \\
\multicolumn{1}{l|}{Geomean}    & \multicolumn{1}{r|}{94.3}          & 95.6                                     & \multicolumn{1}{r|}{\textbf{97.0}}        & \multicolumn{1}{r|}{88.7}          & \multicolumn{1}{r|}{91.5}          & \multicolumn{1}{r|}{87.9}          & 85.5                     \\ \hline
\multicolumn{8}{|c|}{SSIM}                                                                                                                                                                                                                                                                            \\ \hline
\multicolumn{1}{l|}{Isabel}     & \multicolumn{1}{r|}{\textbf{1.00}} & \textbf{1.00}                            & \multicolumn{1}{r|}{\textbf{1.00}}        & \multicolumn{1}{r|}{TO}            & \multicolumn{1}{r|}{\textbf{1.00}} & \multicolumn{1}{r|}{\textbf{1.00}} & \textbf{1.00}            \\
\multicolumn{1}{l|}{Tangaroa}   & \multicolumn{1}{r|}{\textbf{1.00}} & \textbf{1.00}                            & \multicolumn{1}{r|}{\textbf{1.00}}        & \multicolumn{1}{r|}{\textbf{1.00}} & \multicolumn{1}{r|}{\textbf{1.00}} & \multicolumn{1}{r|}{\textbf{1.00}} & 0.99                     \\
\multicolumn{1}{l|}{Earthquake} & \multicolumn{1}{r|}{\textbf{1.00}} & \textbf{1.00}                            & \multicolumn{1}{r|}{TO}                   & \multicolumn{1}{r|}{\textbf{1.00}} & \multicolumn{1}{r|}{\textbf{1.00}} & \multicolumn{1}{r|}{\textbf{1.00}} & 1.00                     \\
\multicolumn{1}{l|}{Ionization} & \multicolumn{1}{r|}{\textbf{1.00}} & \textbf{1.00}                            & \multicolumn{1}{r|}{\textbf{1.00}}        & \multicolumn{1}{r|}{\textbf{1.00}} & \multicolumn{1}{r|}{\textbf{1.00}} & \multicolumn{1}{r|}{\textbf{1.00}} & \textbf{1.00}            \\
\multicolumn{1}{l|}{Miranda}    & \multicolumn{1}{r|}{\textbf{0.34}} & 0.30                                     & \multicolumn{1}{r|}{TO}                   & \multicolumn{1}{r|}{0.30}          & \multicolumn{1}{r|}{\textbf{0.34}} & \multicolumn{1}{r|}{0.30}          & 0.33                     \\
\multicolumn{1}{l|}{S3D}        & \multicolumn{1}{r|}{\textbf{1.00}} & \textbf{1.00}                            & \multicolumn{1}{r|}{TO}                   & \multicolumn{1}{r|}{TO}            & \multicolumn{1}{r|}{\textbf{1.00}} & \multicolumn{1}{r|}{\textbf{1.00}} & 1.00                     \\
\multicolumn{1}{l|}{SCALE}      & \multicolumn{1}{r|}{0.81}          & \textbf{1.00}                            & \multicolumn{1}{r|}{TO}                   & \multicolumn{1}{r|}{TO}            & \multicolumn{1}{r|}{0.86}          & \multicolumn{1}{r|}{0.85}          & 0.76                     \\
\multicolumn{1}{l|}{QMCPACK}    & \multicolumn{1}{r|}{\textbf{1.00}} & \textbf{1.00}                            & \multicolumn{1}{r|}{\textbf{1.00}}        & \multicolumn{1}{r|}{\textbf{1.00}} & \multicolumn{1}{r|}{\textbf{1.00}} & \multicolumn{1}{r|}{\textbf{1.00}} & 1.00                     \\
\multicolumn{1}{l|}{Geomean}    & \multicolumn{1}{r|}{0.85}          & 0.86                                     & \multicolumn{1}{r|}{\textbf{1.00}}        & \multicolumn{1}{r|}{0.78}          & \multicolumn{1}{r|}{0.86}          & \multicolumn{1}{r|}{0.84}          & 0.84                    
\end{tabular}
}
\end{center}
\end{table*}

\section{Summary and Conclusions}
\label{sec:conc_and_future}

This paper describes LOPC, a local-order-preserving data compression algorithm. It is the first lossy compressor that fully preserves the local order (and, by extension, all critical points) of scalar fields. Moreover, it strictly guarantees the user-specified point-wise error bound.

We evaluated LOPC, three state-of-the-art topology-preserving lossy compressors, and five other leading compressors on 2 single- and 6 double-precision inputs. LOPC is the fastest topology-preserving compressor in all cases. At the maximum, it compresses 152,000 times faster than TopoSZ. On average, is 345 times faster than TopoA-augmented SZ3 at an error bound of 1E-4 and a persistence threshold of 1.5E-3. This speed is important for scientific simulations and instruments that produce data at high rates, where the prior state of the art in topology-preserving lossy compression is too slow to be effectively utilized.

LOPC guarantees not only the error bound, which many lossy compressors do not, but also that local ordering and all critical points are preserved. None of the prior topology-preserving compressors can do this without setting the persistence threshold to 0, which slows them down even more.

We hope that LOPC helps enable topology preservation in science where it was previously prohibitively slow to do so.

\section*{Acknowledgments}
\label{sec:acks}
This work has been supported by the U.S.~Department of Energy, Office of Science, Office of Advanced Scientific Research (ASCR), under contract DE-SC0022223.

\bibliographystyle{plain}
\bibliography{references}

\end{document}